\documentclass{article}

\usepackage{arxiv}

\usepackage[utf8]{inputenc} 
\usepackage[T1]{fontenc}    
\usepackage{hyperref}       
\usepackage{url}            
\usepackage{booktabs}       
\usepackage{amsfonts}       
\usepackage{nicefrac}       
\usepackage{microtype}      
\usepackage{lipsum}
\usepackage{graphicx}
\graphicspath{ {./images/} }
\usepackage{amssymb}
\usepackage[caption=false]{subfig}
\usepackage{caption}
\usepackage{multicol}
\usepackage{multirow}
\usepackage{color, colortbl}
\usepackage{graphicx}
\usepackage{float}
\usepackage[usestackEOL]{stackengine}
\usepackage{rotating}
\usepackage{natbib}
\title{Evaluation of individual attributes associated with shared HIV risk behaviors among two network-based studies of people who inject drugs}

\author{
 Valerie Ryan\\
 Department of Computer Science and Statistics\\
 University of Rhode Island\\
 \texttt{vmryan@uri.edu}
  \And
 TingFang Lee, Ashley L. Buchanan \\
  Department of Pharmacy Practice\\
  University of Rhode Island \\
  \texttt{tingfanglee@uri.edu, buchanan@uri.edu} \\
   \And
  Natallia V. Katenka \\
  Department of Computer Science and Statistics\\
  University of Rhode Island
  \texttt{nkatenka@uri.edu} \\
  \And
  Samuel R. Friedman\\
  Department of Population Health\\
  NYU Grossman School of Medicine\\
  \texttt{Samuel.Friedman@nyulangone.org} \\
  \And
  Georgios Nikolopoulos\\
  Medical School, University of Cyprus\\
  \texttt{nikolopoulos.georgios@ucy.ac.cy}
}

\begin{document}
\maketitle
\begin{abstract}
Social context plays an important role in perpetuating or reducing HIV risk behaviors. This study analyzed the network and individual attributes that were associated with the likelihood that people who inject drugs (PWID) will engage in HIV risk behaviors with one another. We analyze data collected in the Social Risk Factors and HIV Risk Study (SFHR) and Transmission Reduction Intervention Project (TRIP) to perform the analysis. Exponential random graph models were used to determine which attributes were associated with the likelihood of people engaging in HIV risk behaviors, such as injection behaviors that are associated with  one another, among PWID. Results across all models and across both data sets indicated that people were more likely to engage in risk behaviors with others who were similar to them in some way (e.g., were the same sex, race/ethnicity, living conditions). In both SFHR and TRIP, we explore the effects of missingness at individual and network levels on the likelihood of individuals to engage in HIV risk behaviors among PWID. In this study, we found that known individual-level risk factors, including housing instability and race/ethnicity, are also important factors in determining the structure of the observed network among PWID. Future development of interventions should consider not only individual risk factors, but communities and social influences leaving individuals vulnerable to HIV risk.
\end{abstract}


\section{Introduction}

According to the Centers for Disease Control and Prevention \citep{cdc2018}, as of 2016, approximately 36.9 million people have HIV worldwide, with 1.8 million new cases reported in 2017 alone.  While the rate of AIDS-related deaths and reports of new HIV cases have decreased since the late 1990s, there has been an increase in HIV diagnoses amongst people who inject drugs (PWID), co-occurring with the rise in opioid and fentanyl use in the United States \citep{fentanyluse2018, frieden2015, burnett2018}.  PWID are an important subpopulation for interventions because they have a higher risk of HIV infection, have higher rates of comorbid diseases and mortality once infected, and have worse health outcomes even when treated with antiretroviral therapy (ART) \citep{altice2010}.

Researchers have evaluated social context and its role in perpetuating or reducing engagement in HIV risk behavior to better deliver HIV interventions among PWID \citep{druginjection2020, friedman2017, latkin1995, unger2006, altice2010}. Two of these observational network studies, the Social Risk Factors and HIV Risk (SFHR) Study and the Transmission Reduction Intervention Project (TRIP), were conducted with participants who engaged in HIV risk behaviors with others. The SFHR Study included participants who had injected drugs within the past year and lived in New York City or the surrounding areas between 1991 and 1993 \citep{friedman2002}. Participants in TRIP were injection drug users and their contacts who lived in Athens, Greece between 2013 and 2015 \citep{nikolopoulos2016}. \cite{dombrowski2013} conducted a study in the SFHR network using exponential random graph models (ERGMs) and examined the influence of transitive closure (i.e., tendency for two people who are connected to someone in common to then connect with each other), and homophily effects (i.e., the tendency of individuals to associate and form connections with others who are similar) for race/ethnicity, age, gender, and number of risk partners on having ties in a network. They found that transitive closure was more important than race/ethnicity, age, gender, and number of injection partners in determining connections in the network \citep{dombrowski2013}; however, they did not discuss other attributes such as education, employment, and living situation. These individual socioeconomic attributes may influence social network formation \citep{nodalattribute2014}, e.g., the position of the individual in the network. Including both demographic and socioeconomic attributes into the model could improve the understanding of the likelihood of people who engage in HIV risk behaviors with one another among PWID.

Although previous studies have identified individual-level risk factors, such as housing instability and race/ethnicity, for engagement in HIV risk behaviors, few studies have evaluated the role of these social and economic risk factors from a network perspective. In this study, we applied ERGMs to data from both the SFHR Study and TRIP  using demographic and socioeconomic individual attributes. We examined network- and individual-level attributes possibly associated with the likelihood of individuals sharing an edge between them defined by engaging in HIV risk behavior together. People are connected if they engage in HIV risk behaviors together, including engaging in unprotected sex, sharing needles or syringes for injection drug use, using drugs around others who also share needles, have sex around others, or have been seen interacting with study participants. 

Most network analysis ignores the problems of missing data by including only individuals with complete attribute data and assuming fully observed network. There are limited number of studies that analyze the effects of missing data with known or unknown missing mechanisms. This study attempts to explore the missingness at individual and network levels, such as missing attributes and unreported risk connections, on the likelihood of individuals to engage in HIV risk behaviors among PWID. We analyzed the SFHR and TRIP using full observed network and only the largest connected component in combination with missing imputation methods for the individual attributes. We consider two missing imputation methods, namely propensity score matching and miss forest imputation. (refs) The findings of this study can be used to understand where and how to target interventions to most effectively decrease engagement in shared HIV risk behaviors among networks of PWID without excluding individuals with only partial information. 

The rest of the paper is organized as follows. We begin by introducing the Network-based studies, SFHR and TRIP, and the statistical methods in Section 2. The analysis results of both network studies will be discussed in Section 3. Section 4 concludes with a discussion. 

\section{Methods}
\subsection{Network-based Studies}

Participants in the Social Risk Factors and HIV Risk (SFHR)  study lived in New York City and the surrounding areas (e.g., New Jersey, New York State, and Connecticut) between 1991 and 1993. Some were recruited by study staff who engaged in ethnographic work where they spent significant amounts of time with people living and using drugs in the neighborhood, while others were brought in by friends or had been given a coupon to participate. Participants were all 18 years of age or older and had injected drugs within the last year \citep{friedman2002}. They completed interviews with study staff and could choose to have their blood drawn and tested for various of diseases, including HIV and Hepatitis B. They provided descriptions of their contacts (people with whom they had sex and/or used drugs, not necessarily in a risky manner). Study staff located those contacts within the neighborhood and asked them to confirm that they knew the participant who had nominated them as a contact. In addition to staff prompted confirmation, some participants also recruited their contacts and brought them to the storefront for confirmation. Researchers also collected data on demographic information (e.g., age, race, ethnicity, sex, employment status), health attitudes and beliefs, and health-seeking behavior. As a result, 767 people enrolled in the study and 516 shared connections between those people \citep{friedman2002}.

Participants in the Transmission Reduction Intervention Project (TRIP) study were people who inject drugs and their contacts who lived in Athens, Greece, between 2013 and 2015. Those who were initially recruited into the study were injection drug users referred to the study by HIV testing centers in Athens. These participants were initially recruited by ARISTOTLE, a program that followed a respondent-driven sampling (RDS) design and sought to contract trace PWID and enroll them in HIV care \citep{aristotle}. For individuals initially recruited into the study, it was determined whether they were recently infected with HIV (within the last 6 months) or were considered long-term HIV infected (more than 6 months prior)\citep{nikolopoulos2016}. Researchers then used two-wave contact tracing for each initial recruit, asking them about their sexual and drug use partners. Once those nominated participants were contacted and agreed to participate, they had their HIV status ascertained. Those who were HIV positive were tested using the Limited Antigen Avidity (LAg) assay to determine if they were recently infected with HIV ($<$ 130 days since infection) or long-term infected ($>$ 130 days since infection) \citep{nikolopoulos2016}. It is important to differentiate between short-term and long-term infected because those who are recently infected are more contagious and likely not to have many symptoms while in their most acutely infectious period \citep{selik2018, volz2013}. Both those who were recently infected and those who were long-term infected were asked to provide contact information about their sexual and drug use partners. Many of those individuals were recruited into the study \citep{nikolopoulos2016}.

In TRIP, participants completed computer-assisted and paper interviews, and also had their HIV status ascertained. If HIV status was positive,  the recency of infection was determined  with the LAg test. Participants provided demographic information and their contacts’ information, answered questions about engagement in risk behaviors, their HIV status, substance use, access to care, HIV knowledge, stigma, injection norms, and their opinions on the project. Follow-up interviews were conducted with participants about six months after they completed their baseline interview. The data were collected over two years of follow-up with 356 people enrolled in the study and has 542 shared connections. \citep{nikolopoulos2016}.

\subsection{Statistical Methods}
This study used existing, de-identified data from two observational network-based studies, namely, SFHR and TRIP. Each node in the network represents a participant in the study. If the participants engage in HIV risk behavior together, then there exists an edge between the corresponding nodes. Descriptive statistics for the participants' attributes are reported for each study network in Tables \ref{tab:SHFRdescriptivestat}, \ref{tab:TRIPdescriptivestat}, and \ref{tab:network}. The primary attributes in SFHR were age, sex, race/ethnicity, education level, employment, living situation, and marital status. These variables were chosen because they could be important for the delivery of interventions and are easy to ascertain \citep{ware1981}. The primary attributes of interest in the TRIP analyses were age, sex, nationality, education level, employment, and living situation. Sexual identity was not included as 97\% of participants identified as heterosexual. The detailed description of individual attributes and data pre-processing are described in Appendix A. 

\begin{table}
  \centering
  \caption{Descriptive statistics of SFHR attribute variables. These include counts and frequencies for each variable in both the full network, which includes the counts and percentages of all nodes, and the largest connected component (LCC), as well as means and ranges for age, a continuous variable. }
    \begin{tabular}{llrrrr}
    \midrule
          &       & \multicolumn{2}{c}{Full Network } & \multicolumn{2}{c}{LCC} \\
          &       & \multicolumn{2}{c}{N=767} & \multicolumn{2}{c}{N=277} \\
          &       & {N} & {\%} & \multicolumn{1}{c}{N} & \multicolumn{1}{c}{ \%} \\
    \midrule
    Sex   & Male  & 541   & 71    & 195   & 70 \\
    \midrule
    \multirow{4}[2]{*}{Race} & African  & 206   & 27    & 81    & 29 \\
          & American Latinx & 311   & 41    & 87    & 31 \\
          & White & 243   & 32    & 104   & 38 \\
          & Other & 7     & 1     & 5     & 2 \\
    \midrule
    \multirow{2}[2]{*}{Education} & {Less than high school} & 472   & 61    & 176   & 64 \\
          & {High school or more} & 295   & 39    & 101   & 36 \\
    \midrule
    \multirow{2}[2]{*}{Employment} & Employed & 78    & 10    & 24    & 9 \\
          & Unemployed & 689   & 90    & 253   & 91 \\
    \midrule
    \multicolumn{1}{l}{\multirow{4}[2]{*}{\shortstack{Living\\Situation}}} & In own place & 244   & 32    & 76    & 27 \\
          & Someone else's place & 370   & 48    & 132   & 48 \\
          & Homeless & 137   & 18    & 62    & 22 \\
          & Missing & 16    & 2     & 7     & 3 \\
    \midrule
    \multicolumn{1}{l}{\multirow{3}[1]{*}{\shortstack{Marital\\Status}}} & Single & 400   & 51    & 136   & 49 \\
          & Married & 167   & 22    & 53    & 19 \\
          & Divorced & 200   & 26    & 88    & 32 \\
   \midrule
    Age   & Mean and SD & 35.25 & 6.97  & 35.19 & 6.63 \\
    \bottomrule
    \end{tabular}%
  \label{tab:SHFRdescriptivestat}%
\end{table}%

\begin{table}
  \centering
  \caption{Descriptive statistics of TRIP attribute variables. These include counts and frequencies for each variable in both the full network, which includes all nodes, and the largest connected component (LCC), as well as means and standard deviations for age, a continuous variable.}
    \begin{tabular}{llrrrr}
    \midrule
          &       & \multicolumn{2}{c}{Full Network } & \multicolumn{2}{c}{LCC} \\
          &       & \multicolumn{2}{c}{N=356} & \multicolumn{2}{c}{N=241} \\
          &       & \multicolumn{1}{c}{N} & \multicolumn{1}{c}{\%} & \multicolumn{1}{c}{N} & \multicolumn{1}{c}{ \%} \\
    \midrule
    Sex   & Male  & 281   & 79    & 191   & 79 \\
    \midrule
    Nationality & Greek & 323   & 91    & 212   & 88 \\
    \midrule
    \multirow{3}[2]{*}{Education} & Primary School & 113   & 32    & 74    & 31 \\
          & High School & 196   & 55    & 135   & 56 \\
          & Post High School & 47    & 13    & 32    & 13 \\
    \midrule
    \multirow{4}[2]{*}{Employment} & Employed & 61    & 17    & 38    & 16 \\
          & Unemployed: looking for work & 89    & 25    & 51    & 21 \\
          & Can't work, health reasons & 161   & 45    & 117   & 49 \\
          & Other & 45    & 13    & 35    & 14 \\
    \midrule
    \multicolumn{1}{l}{\multirow{4}[2]{*}{\shortstack{Living\\Situation}}} & Paying rent & 75    & 21    & 39    & 16 \\
          & Not paying rent & 193   & 54    & 124   & 51 \\
          & Homeless & 80    & 23    & 71    & 29 \\
          & Missing & 8     & 2     & 7     & 4 \\
    \midrule
    Age   & (Mean and SD) & 35.87 & 8.39  & 35.99 & 8.54 \\
    \bottomrule
    \end{tabular}%
  \label{tab:TRIPdescriptivestat}%
\end{table}%

\begin{table}
  \begin{center}
  \footnotesize
  \caption{Network descriptive statistics of SFHR (left) and TRIP (right), calculated for both the full network and the largest connected component (LCC).}
    \begin{tabular}{lrrrr}
          & \multicolumn{2}{c}{SFHR} & \multicolumn{2}{c}{TRIP} \\
          & \multicolumn{1}{c}{Full Network} & \multicolumn{1}{c}{LCC} & \multicolumn{1}{c}{Full Network} & \multicolumn{1}{c}{LCC} \\
    \midrule
    Node Count & 767   & 277   & 356   & 241 \\
    Edge Count & 516   & 380   & 542   & 502 \\
    Assortativity & 0.1   & -0.0004 & 0.2   & 0.15 \\
    Transitivity & 0.12  & 0.11  & 0.24  & 0.23 \\
    Average degree (SD) & 1.35 (2.25) & 2.74 (3.17) & 3.04 (3.46) & 4.17 (3.6) \\
    Average betweenness centrality & 0.0008 & 0.017 & 0.005 & 0.016 \\
    Density & 0.002 & 0.01  & 0.009 & 0.017 \\
    \bottomrule
    \end{tabular}%
    \label{tab:network}%
     \end{center}
    {\footnotesize \flushleft Note: transitivity, average betweenness centrality, and density range from 0 (low) to 1 (high); average degree ranges from 0 to the number of people in a network minus 1; and assortativity ranges from $-1$ to $+1$, indicating negatively related to positively related.}
\end{table}%

ERGMs were used to determine which network and individual attributes were associated with individuals possibly engaging in HIV risk behavior with others in both the SFHR and TRIP networks. Individual attributes examined in the current study, including sex, employment status, age, education level, living situation, race/ethnicity, marital status, and nationality, were used in the ERGMs. All nodal attribute variables were included using three different terms: node match, node factor, and node mix. Three separate models were used for nodal attribute variables in each study network. The first  model included the effects of node match (the odds of people being connected based on a shared attribute). The second model included the effects of node factor (comparing the odds of people being connected across levels of an attribute variable). And the third model included the effects of node mix (comparing the odds of people being connected within and between levels of an attribute variable). These were modeled separately because the effects of every term for each attribute could not be estimated when they were included together in a single model. After node match, node factor, and node mix models were run separately. The final models for each study were created containing a combination of those terms, based on which terms had been statistically significant in the previous models. First, univariate models were developed that only included terms for each attribute separately. If a term was significant at the $p < 0.2$ level in the univariate models, the term was included in the adjusted models. The entire network sample (full network) and the largest connected component were modeled separately, and the parameter estimates were compared to see if there are differences in the results. The full network model included all participants, some of whom were isolates (they did not share any edges with others in the network). The evaluation of the goodness-of-fit of all the models indicated no evidence of a lack of fit.

We also incorporated network structure in the ERGMs. The only network structure term that could consistently be estimated across all scenarios was edges. The only network term, apart from edges, that could be included in some models was the geometrically-weighted degree term, and this could only be estimated in models of the largest connected component in the SFHR data set. Geometrically-weighted degree can be thought of as a measure of “popularity,” meaning it estimates the likelihood of a person having a tie in the network based on their degree, which is the number of connections (edges) they have in the network and the degree of others \citep{hunter2008}.

Missing attribute information for the nodes is common in network-based studies, particularly among vulnerable populations. In the presence of missing nodal attribute data, techniques to address missing data in the analysis are needed, as ERGMs do not converge when missing data are included in the model. In network studies, excluding a person with missing data could significantly change network structure by removing their ties with other people in the network \citep{gile2006}. To address missing nodal attribute data, we used two missing data imputation methods: propensity score matching and multiple imputation using missForest. Propensity score matching requires the propensity score model to be correctly specified \citep{dagostino2000, little2002}. The miss forest non-parametric technique uses observed information to predict missing values, but does not require any modeling assumptions \citep{stekhoven2012}. For both methods, data were imputed using the same set of attributes that were candidates for the ERGMs. For the TRIP data, propensity scores and missForest were generated to impute missing data for the variable about where a person was currently living based on sex, employment status, nationality, education, and age. For the SFHR data, propensity scores and missForest were generated to impute missing data for the variable about where a person was currently living based on sex, race/ethnicity, employment status, marital status, education, and age. This secondary data analysis study was reviewed and approved by the Institutional Review Board (IRB) at the University of Rhode Island.

\section{Results}
The full networks and largest connected components (LCC) of SFHR and TRIP were analyzed using complete cases and two missing data imputation techniques, propensity score matching and miss forest. The results across full networks and LCC using complete cases and both missing data imputation methods are comparable. In this section, we focus the discussion on the results of the final models on the full network of SFHR and TRIP using complete cases. The model results using complete cases can be found in Appendix B.

\subsection{SFHR}
The final model using complete cases for SFHR full network included terms for node mixing based on living situation and race/ethnicity, homophily terms for sex and marital status, and a node factor term for educational attainment (Table B4). Those who lived in their own place were 65\% less likely to be connected to others who lived in their own place (odds ratio (OR) = 0.35, 95\% confidence interval (CI) = (0.23, 0.52)), 73\% less likely to connect with those who lived in someone else’s place (OR = 0.27 , 95$\%$ CI = (0.23, 0.52)), and similarly, 73\% less likely to be connected to those who were homeless (OR = 0.27 , 95$\%$ CI = (0.18, 0.41)), compared to ties between two people who were homeless. Those who lived in someone else’s place were 67\% less likely to be connected to others who lived in someone else’s place (OR = 0.33, 95\% CI = (0.23, 0.46)), and those who were homeless (OR = 0.4 , 95$\%$ CI = (0.28, 0.56)), compared to ties between two people who were homeless. 

We also observed homophily by race, sex, and marital status. African Americans were 1.44 times (95\% CI = (1.11, 1.87)) more likely to be connected to other African Americans, compared to the likelihood of a connection between two White people in the network. Males had 1.27 times (95\% CI = (1.05, 1.54)) the odds of connecting with other males. Females had 1.55 times (95\% CI = (1.13, 2.11)) the odds of connecting with other females. Single people had 1.44 times (95\% CI = (1.19, 1.76)) the odds of being connected to other single people. Lastly, people with a high school education or more were 23\% less likely to have ties in the network, compared to those with less than high school education. 

The two ERGM models using geometrically-weighted degree term were the node match model and node factor model (Table B5). The node match models with and without the geometrically-weighted degree were similar; however, being in a relationship was statistically significant in this model. Those who were in a relationship having 1.72 times (95\% CI = (1.03, 2.86)) the odds of being connected with others who were also in relationships in the network. The node factor models with and without the geometrically-weighted degree were again similar. However, in the model with the geometrically-weighted degree term a high school education or more was no longer a statistically significant term. The geometrically-weighted degree term was statistically significant in both node match and node factor models, indicating a potential popularity effect, according to which people were more likely to form ties with higher-degree people in the network.

\subsection{TRIP}
The final model for the TRIP full network using complete cases included node mixing terms based on living situation, sex and nationality, and node factor terms based on educational attainment and employment status (Table B9). Male to male (OR = 0.61, 95\% CI = (0.43, 0.86)) and male to female (OR = 0.59, 95\% CI = (0.41, 0.84)) connections were significantly less likely to occur, compared to female to female connections. Connections between those who were Greek and those who were not (OR = 0.34, 95\% CI = (0.20, 0.56)) and connections between two Greek people (OR = 0.43, 95\% CI = (0.26, 0.72)) were also less likely to occur, compared to connections between two non-Greek people. Those who paid rent were less likely to be connected to others who paid rent (OR = 0.21, 95\% CI = (0.12, 0.36)), those who did not pay rent (OR = 0.18, 95\% CI = (0.13, 0.25)), and those who were homeless (OR = 0.23, 95\% CI = (0.16, 0.34)), compared to connections between two people who were homeless. Those who did not pay rent were less likely to be connected to those who did not pay rent (OR = 0.25, 95\% CI = (0.19, 0.34)), and those who were homeless (OR = 0.32, 95\% CI = (0.25, 0.42)), compared to connections between two people who were homeless. People with post high school education were 1.35 times (95\% CI = (1.12, 1.63)) more likely to have ties in the network, compared to those with less than a high school education. People who could not work for health reasons were 1.34 times (95\% CI = (1.08, 1.66)) more likely to have ties in the network, compared to those who were employed. People who marked “other” as their employment status had 1.31 times (95\% CI = (1.02, 1.70)) the odds of having connections in the network, compared to those who were employed.

\section{Discussion}
In this paper, we analyzed the network attributes and individual attributes associated with the likelihood of people engaging in HIV risk behaviors with each other among PWID in two network-based studies SFHR and TRIP. People who consistently had the highest odds of being connected within both networks across all models were those experiencing housing instability. They were more likely to have network connections and were more likely to be connected with one another. Although often challenging to reach and sustain engagement with public health interventions, these individuals represent a subpopulation that has the potential to benefit from interventions that leverage network connections due to their positioning in the observed network. Interventions for PWID who experience homelessness could include establishing safe injection sites or outreach to deliver harm reduction interventions by visiting places they frequent or at community centers, which could decrease risky injection drug use and overdose deaths. People who experience housing instability are a particularly vulnerable subpopulation, and often lack access to medical care, treatment services, and other supports, so the interventions could be more readily accessible to them through both outreach efforts and from within their communities using intraventions. \citep{intravention2004}.

The results suggest that study participants tend to engage in potentially risky behavior with others that are similar based on sex and shared race/ethnicity or nationality. Delivering interventions to those on the periphery in the network, such as those employed and live in their own place/pay rent, could be challenging because the peer influence and reinforcement may have less of an impact. Considering the attributes of those engaging in risk behavior with each other could better inform the development of interventions. For example, an intervention is delivered by recruiting individuals and expecting them to share information about the intervention and recruit others they know. Researchers must ensure that they are intervening in diverse groups of individuals. Interventions aimed at HIV prevention among PWID should focus on increasing accessibility to as many people as possible. Network analysis is one possible approach to determine individuals and groups that might be good candidates for interventions, particularly when there is a social component delivering an intervention.
 
The study has potential limitations. The missing data imputation techniques, propensity score matching and missForest, assume independent observations. In future work, the inclusion of network structure variables in missing attribute data imputation techniques is important to align methodology with the observed data structure. In alternative data sources, ERGMs may be able to include additional network structure variables to examine whether there are any structural effects on engagement in potentially risky behavior. Conducting future longitudinal studies among PWID now would also be very informative to see how behaviors and network structures have changed over time. There has been an increase in heroin use and prescription opioid use, specifically fentanyl, coinciding with a sharp rise in opioid overdoses in recent years. Development intervention strategies according to the information from longitudinal studies among PWID can be of benefit to mitigate opioid overdose. 

\section*{Acknowledgement}
These findings are presented on behalf of Social Risk Factors and HIV Risk Study (SFHR) and the Transmission Reduction Intervention Project (TRIP). We would like to thank all of the SFHR and TRIP investigators, data management teams, and participants who contributed to this project. The project described was supported grant 1DP2DA046856-01 by Avenir Award Program for Research on Substance Abuse and HIV/AIDS (DP2) from National Institute on Drug Abuse of the National Institutes of Health and grant P30 DA011041 Ending HIV/AIDS among People Who Use Drugs: Overcoming Challenges by Center for Drug Use and HIV Research. The content is solely the responsibility of the authors and does not necessarily represent the official views of the National Institutes of Health.

\appendix

\section{Variable Definitions}

\subsection{Social Risk Factors and HIV Risk Study (SFHR) }
 Age was calculated as the difference between the person's birthday and their date of study enrollment; it is a continuous variable. Sex was male or female. Race was grouped as African American, Latinx, White, and Others. The first three groups already existed in the data; the Others category included those who identify as ‘Asian/Pacific Islander’, ‘Native American’, or ‘Other.’ The ‘Other’ category was small ($n = 7$) and was thus combined with the African American category for analyses. Education was dichotomized as less than high school or high school or more. The less than high school category combined those who endorsed completing ‘Elementary’ or ‘Less than high school graduation/12th grade’ as their highest education level. The high school or more combined ‘high school graduation’, ‘some college’, and ‘college graduation.’

Work was dichotomized into unemployed and employed, with unemployed combining ‘unemployed’ and ‘unable to work – disabled.’ The employed category included ‘regular full-time work’, ‘regular part-time work’, ‘occasional work’, ‘self-employed’, ‘retired’, ‘student’, and ‘homemaker’. Living situation was reduced to three categories: ‘homeless’, ‘in own apartment/house’, or ‘someone else’s apartment/house.’ The homeless category included those who lived ‘on the streets’, ‘in an abandoned building’, ‘in a car/van/truck’, ‘in a subway train or station’, and ‘in a shelter or welfare boarding home.’ The other two living situation categories, ‘in own apartment/house’ and ‘in someone else’s apartment/house’ were categories already existing in the data. Those who answered ‘other’ to the question are categorized as having missing data for this question. Marital status was grouped as single, in a relationship, or divorced. The single category included those who were ‘never married’. The married category included those who were ‘married’, those who were ‘widowed’, and those who were ‘informally married/living together’. The divorced category included those who were ‘divorced’ and ‘separated.’

\subsection{Transmission Reduction Intervention Project (TRIP)}
 Age, a continuous variable, was calculated as the difference between a person’s birthday and their study enrollment date. Participants were asked if they were male, female, or transgender. No participants endorsed being transgender, so only male and female were used. Nationality was split into Greek or not Greek, as the large majority of participants were Greek. The 7 original groups of education level were collapsed into 3 categories: completed at most primary school (combining ‘less than primary school’ and ‘primary school or similar’), completed at most high school (combining ‘first 3 years of high school’ and ‘last 3 years of high school’), and completed some education beyond high school (combining ‘vocational training institutes or private universities/colleges’, ‘public technical education institutes or universities’, and ‘postgraduate studies/PhD’). 

There were 9 categories for employment, combined into 4 categories: employed (‘employed full-time’, ‘employed part-time’, ‘run my own business’, ‘have occasional earnings’, ‘homemaker’, and ‘retired’), unemployed but looking for work, can’t work for health reasons, and other. Living situation had 8 categories, which were grouped into 4 categories: paying rent (‘in your own house/apartment’, ‘in your family/relative’s /friend’s house paying rent’, and ‘in a rented house either renting alone or with a partner or friend’), not paying rent (‘in your family/relative’s/friend’s house not paying rent’ and ‘more than one of the previous categories’, which included not paying rent), homeless, and missing (combining ‘other’ and ‘not asked’).

\section{SFHR and TRIP Model results}

This section includes all model results on SFHR and TRIP networks. Three separate models across complete cases analysis and missing data imputation using propensity score matching and miss forest were used for each network; one model included the effects of node match (the odds of people being connected based on a shared attributes), one included the effects of node factor (comparing the odds of people being connected across levels of an attribute variable), and the last included the effects of node mix (comparing the odds of people being connected within and between levels of an attribute variable). These were modeled separately because the effects of every term for each attribute could not be estimated when they were included together in one model. After node match, node factor, and node mix models were run separately, final models were created containing a combination of those terms, based on which terms had been statistically significant in the previous models.

The only network term, apart from edges, that could be included in models was the geometrically-weighted degree term and this could only be estimated in models of the largest connected component in the SFHR data set.

\begin{sidewaystable}\setlength{\tabcolsep}{1.8pt}
\begin{table}[H]
  \begin{center}\footnotesize
  \caption{ERGM node match model results in SFHR for complete cases and both imputation methods, modeled in both the full network and the largest connected component (LCC).}
    \begin{tabular}{clcccccccccccc}
          & \multicolumn{1}{r}{} & \multicolumn{4}{c}{Complete Case} & \multicolumn{4}{c}{Propensity Score Matching} & \multicolumn{4}{c}{Miss Forest} \\
          & \multicolumn{1}{r}{} & \multicolumn{2}{c}{Full Network} & \multicolumn{2}{c}{LCC} & \multicolumn{2}{c}{Full Network} & \multicolumn{2}{c}{LCC} & \multicolumn{2}{c}{Full Network} & \multicolumn{2}{c}{LCC} \\
          & \multicolumn{1}{r}{} & {OR} & 95\% CI & {OR} & 95\% CI & {OR} & 95\% CI & {OR} & 95\% CI & {OR} & 95\% CI & {OR} & 95\% CI \\
    \midrule
    {\shortstack{Network\\Characteristic}} & {\shortstack{Edges\\\textcolor{white}{Edges}}} & {0.0004$^\ddagger$} & \shortstack{(0.0003,\\0.0005)} & {0.002$^\ddagger$} & \shortstack{(0.0016,\\0.003)} & {0.0004$^\ddagger$} & \shortstack{(0.0003,\\0.0005)} & {0.002$^\ddagger$} & \shortstack{(0.002,\\0.003)} & {0.0004$^\ddagger$} & \shortstack{(0.0003,\\0.0005)} & {0.002$^\ddagger$} & \shortstack{(0.002,\\0.003)} \\
    \midrule
    \multicolumn{1}{c}{\multirow{2}[2]{*}{Sex}} & Male  & {1.29$^\dagger$} & (1.06, 1.56) & {1.54$^\ddagger$} & (1.22, 1.95) & {1.26*} & (1.04, 1.52) & {1.53$^\ddagger$} & (1.22, 1.92) & {1.26*} & (1.04, 1.52) & {1.53$^\ddagger$} & (1.22, 1.92) \\
          & Female & {1.52$^\dagger$} & (1.12, 2.08) & {1.72$^\dagger$} & (1.18, 2.51) & {1.56$^\dagger$} & (1.16, 2.11) & {1.76$^\dagger$} & (1.23, 2.52) & {1.56$^\dagger$} & (1.16, 2.11) & {1.76$^\dagger$} & (1.23, 2.51) \\
    \midrule
    \multicolumn{1}{c}{\multirow{3}[2]{*}{Race}} & \shortstack{African\\American} & {6.72$^\ddagger$} & (5.28, 8.55) & {4.44$^\ddagger$} & (3.36, 5.85) & {6.58$^\ddagger$} & (5.18, 8.35) & {4.45$^\ddagger$} & (3.38, 5.86) & {6.57$^\ddagger$} & (5.17, 8.35) & {4.44$^\ddagger$} & (3.37, 5.84) \\
          & Latinx & {2.71$^\ddagger$} & (2.12, 3.46) & {3.65$^\ddagger$} & (2.73, 4.88) & {2.72$^\ddagger$} & (2.15, 3.46) & {3.48$^\ddagger$} & (2.62, 4.62) & {2.73$^\ddagger$} & (2.15, 3.46) & {3.50$^\ddagger$} & (2.63, 4.64) \\
          & White & {4.77$^\ddagger$} & (3.74, 6.10) & {2.77$^\ddagger$} & (2.07, 3.70) & {4.81$^\ddagger$} & (3.79, 6.11) & {2.84$^\ddagger$} & (2.14, 3.77) & {4.81$^\ddagger$} & (3.78, 6.11) & {2.84$^\ddagger$} & (2.14, 3.76) \\
    \midrule
    \multicolumn{1}{c}{\multirow{2}[2]{*}{Education}} & \shortstack{Less than\\high school} & {1.57$^\ddagger$} & (1.29, 1.90) & {1.39$^\dagger$} & (1.11, 1.74) & {1.53$^\ddagger$} & (1.27, 1.85) & {1.39$^\dagger$} & (1.11, 1.73) & {1.53$^\ddagger$} & (1.26, 1.84) & {1.39$^\dagger$} & (1.11, 1.74) \\
          & \shortstack{High School\\or more} & 1.02  & (0.76, 1.35) & 1.07  & (0.76, 1.52) & 0.97  & (0.73, 1.29) & 1.05  & (0.75, 1.48) & 0.97  & (0.73, 1.28) & 1.05  & (0.75, 1.48) \\
    \midrule
    \multicolumn{1}{c}{\multirow{2}[2]{*}{Employment}} & Employed & 1.59  & (0.69, 3.66) & 1.5   & (0.46, 4.88) & 1.58  & (0.69, 3.65) & 1.48  & (0.46, 4.80) & 1.58  & (0.69, 3.64) & 1.48  & (0.46, 4.81) \\
          & Unemployed & 1.27  & (0.98, 1.63) & 1.15  & (0.85, 1.56) & 1.28  & (0.998, 1.65) & 1.16  & (0.86, 1.57) & {1.29*} & (1.00, 1.65) & 1.16  & (0.86, 1.57) \\
    \midrule
    \multicolumn{1}{c}{\multirow{3}[2]{*}{\shortstack{Living\\Situation}}} & In own place & 1.14  & (0.84, 1.55) & 1.24  & (0.82, 1.87) & 1.11  & (0.82, 1.49) & 1.22  & (0.82, 1.81) & 1.15  & (0.85, 1.56) & 1.27  & (0.85, 1.91) \\
          & \shortstack{Someone\\else's place} & 1.05  & (0.85, 1.31) & 1.04  & (0.81, 1.35) & 1.04  & (0.84, 1.28) & 1.03  & (0.80, 1.33) & 1.05  & (0.85, 1.29) & 1.03  & (0.81, 1.32) \\
          & Homeless & {3.17$^\ddagger$} & (2.33, 4.30) & {2.89$^\ddagger$} & (2.08, 4.01) & {3.12$^\ddagger$} & (2.31, 4.22) & {2.82$^\ddagger$} & (2.04, 3.90) & {3.18$^\ddagger$} & (2.35, 4.32) & {2.93$^\ddagger$} & (2.11, 4.05) \\
    \midrule
    \multicolumn{1}{c}{\multirow{3}[2]{*}{\shortstack{Marital\\Status}}} & Single & {1.46$^\ddagger$} & (1.20, 1.78) & {1.83$^\ddagger$} & (1.45, 2.31) & {1.42$^\ddagger$} & (1.17, 1.72) & {1.84$^\ddagger$} & (1.47, 2.31) & {1.43$^\ddagger$} & (1.18, 1.73) & {1.84$^\ddagger$} & (1.46, 2.30) \\
          & Married & 1.46  & (0.97, 2.19) & 1.63  & (0.95, 2.81) & 1.46  & (0.98, 2.16) & 1.64  & (0.98, 2.74) & 1.46  & (0.98, 2.16) & 1.64  & (0.98, 2.73) \\
          & Divorced & 1.27  & (0.91, 1.78) & 1.11  & (0.77, 1.62) & 1.28  & (0.92, 1.78) & 1.09  & (0.76, 1.58) & 1.28  & (0.92, 1.78) & 1.08  & (0.75, 1.57) \\
    \bottomrule
    \end{tabular}%
          
  \end{center}
  \label{tab:addlabel}%
  \flushleft \footnotesize{Note: the node match model represents the odds of two people being connected in a network based on sharing some attributes, e.g., males being connected to other males in the network.\\
OR = odds ratio\\
95\% CI is the 95\% confidence interval of the odds ratio
* indicates $p < 0.05$, $\dagger$ indicates $p < 0.01$, and $\ddagger$ indicates $p < 0.001$}
\end{table}%
\end{sidewaystable}

\begin{sidewaystable}\setlength{\tabcolsep}{1.8pt}
\begin{table}[H]
  \begin{center}\footnotesize
  \caption{ERGM node factor model results in SFHR for complete cases and both imputation methods, modeled in both the full network and the largest connected component (LCC).}
    \begin{tabular}{clcccccccccccc}
          & \multicolumn{1}{l}{} & \multicolumn{4}{c}{Complete Case} & \multicolumn{4}{c}{Propensity Score Matching} & \multicolumn{4}{c}{Miss Forest} \\
          & \multicolumn{1}{l}{} & \multicolumn{2}{c}{Full Network} & \multicolumn{2}{c}{LCC} & \multicolumn{2}{c}{Full Network} & \multicolumn{2}{c}{LCC} & \multicolumn{2}{c}{Full Network} & \multicolumn{2}{c}{LCC} \\
          & \multicolumn{1}{l}{} & OR    & \multicolumn{1}{c}{95\% CI} & \multicolumn{1}{c}{OR} & \multicolumn{1}{c}{95\% CI} & \multicolumn{1}{c}{OR} & \multicolumn{1}{c}{95\% CI} & \multicolumn{1}{c}{OR} & \multicolumn{1}{c}{95\% CI} & \multicolumn{1}{c}{OR} & \multicolumn{1}{c}{95\% CI} & \multicolumn{1}{c}{OR} & \multicolumn{1}{c}{95\% CI} \\
    \midrule
    \shortstack{Network\\Characteristic} & {\shortstack{Edges\\\textcolor{white}{Edges}}} & 0.002$^\ddagger$ & {(0.001, 0.003)} & {0.01$^\ddagger$} & {(0.007, 0.02)} & {0.002$^\ddagger$} & {(0.001, 0.003)} & {0.01$^\ddagger$} & {(0.006, 0.02)} & {0.002$^\ddagger$} & {(0.001, 0.003)} & {0.01$^\ddagger$} & {(0.006, 0.02)} \\
    \midrule
    \multicolumn{1}{c}{\multirow{2}[2]{*}{Sex}} & Female & \multicolumn{1}{c}{0.96} & {(0.84, 1.11)} & 0.87  & {(0.73, 1.03)} & 0.99  & {(0.87, 1.14)} & 0.9   & {(0.76, 1.06)} & 0.99  & {(0.87, 1.14)} & 0.89  & {(0.76, 1.05)} \\
          & Male  & \multicolumn{12}{c}{Reference Level} \\
    \midrule
    \multicolumn{1}{c}{\multirow{3}[2]{*}{Race}} & Latinx & 0.71$^\ddagger$ & {(0.61, 0.83)} & 0.99  & {(0.83, 1.18)} & {0.72$^\ddagger$} & {(0.62, 0.84)} & 0.96  & {(0.81, 1.15)} & {0.73$^\ddagger$} & {(0.62, 0.85)} & 0.97  & {(0.81, 1.16)} \\
          & White & \multicolumn{1}{c}{0.86} & {(0.73, 1.00)} & {0.77$^\dagger$} & {(0.64, 0.92)} & 0.88  & {(0.75, 1.03)} & {0.79$^\dagger$} & {(0.66, 0.94)} & 0.88  & {(0.76, 1.03)} & {0.79$^\dagger$} & {(0.66, 0.94)} \\
          & \shortstack{African\\American} & \multicolumn{12}{c}{Reference Level} \\
    \midrule
    \multicolumn{1}{c}{\multirow{2}[2]{*}{Education}} & \shortstack{High School\\or more} & 0.77$^\ddagger$ & {((0.67, 0.88)} & {0.85* } & {(0.72, 0.996)} & {0.77$^\ddagger$} & {(0.67, 0.88)} & {0.85*} & {(0.72, 0.99)} & {0.77$^\ddagger$} & {(0.67, 0.88)} & {0.84*} & {(0.72, 0.99)} \\
          & \shortstack{Less than\\high school} & \multicolumn{12}{c}{Reference Level} \\
    \midrule
    \multicolumn{1}{c}{\multirow{2}[2]{*}{Employment}} & Unemployed & \multicolumn{1}{c}{1.15} & {(0.92, 1.45)} & 1.07  & {(0.81, 1.42)} & 1.17  & {(0.93, 1.46)} & 1.08  & {(0.82, 1.42)} & 1.18  & {(0.94, 1.47)} & 1.08  & {(0.82, 1.42)} \\
          & Employed & \multicolumn{12}{c}{Reference Level} \\
    \midrule
    \multicolumn{1}{c}{\multirow{3}[2]{*}{\shortstack{Living\\Situation}}} & \shortstack{Someone\\else's place} & \multicolumn{1}{c}{1.1} & {(0.94, 1.29)} & 1.08  & {(0.88, 1.31)} & 1.11  & {(0.95, 1.29)} & 1.08  & {(0.89, 1.31)} & 1.1   & {(0.95, 1.29)} & 1.06  & {(0.87, 1.29)} \\
          & Homeless & 1.60$^\ddagger$ & {(1.34, 1.91)} & {1.54$^\ddagger$} & {(1.24, 1.91)} & {1.60$^\ddagger$} & {(1.35, 1.91)} & {1.52$^\ddagger$} & {(1.23, 1.87)} & {1.62$^\ddagger$} & {(1.35, 1.93)} & {1.53$^\ddagger$} & {(1.24, 1.90)} \\
          & In own place & \multicolumn{12}{c}{Reference Level} \\
    \midrule
    \multicolumn{1}{c}{\multirow{3}[2]{*}{\shortstack{Marital\\Status}}} & Married & \multicolumn{1}{c}{0.95} & {(0.80, 1.13)} & 0.87  & {(0.70, 1.08)} & 0.96  & {(0.81, 1.13)} & 0.83  & {(0.67, 1.03)} & 0.97  & {(0.82, 1.14)} & 0.84  & {(0.68, 1.04)} \\
          & Divorced & \multicolumn{1}{c}{0.88} & {(0.76, 1.03)} & {0.72$^\ddagger$} & {(0.61, 0.86)} & 0.91  & {(0.78, 1.05)} & {0.72$^\ddagger$} & {(0.61, 0.86)} & 0.91  & {(0.78, 1.05)} & {0.72$^\ddagger$} & {(0.60, 0.86)} \\
          & Single & \multicolumn{12}{c}{Reference Level} \\
    \bottomrule
    \end{tabular}%
  \end{center}
  \label{tab:addlabel}%
    \flushleft \footnotesize{Note: the node factor model is used to compare people across levels of an attribute variable to see if people are more or less likely to have ties in the network, compared to a reference level.\\
OR = odds ratio\\
95\% CI is the 95\% confidence interval of the odds ratio
* indicates $p < 0.05$, $\dagger$ indicates $p < 0.01$, and $\ddagger$ indicates $p < 0.001$}
\end{table}%
\end{sidewaystable}

\begin{sidewaystable}\setlength{\tabcolsep}{1.8pt}
\begin{table}[H]
  \begin{center}\scriptsize
  \caption{ERGM node mix model results in SFHR for complete cases and both imputation methods, modeled in both the full network and the largest connected component (LCC).}
    \begin{tabular}{clcccccccccccc}
          & \multicolumn{1}{l}{} & \multicolumn{4}{c}{Complete Case} & \multicolumn{4}{c}{Propensity Score Matching} & \multicolumn{4}{c}{Miss Forest} \\
          & \multicolumn{1}{l}{} & \multicolumn{2}{c}{Full Network} & \multicolumn{2}{c}{LCC} & \multicolumn{2}{c}{Full Network} & \multicolumn{2}{c}{LCC} & \multicolumn{2}{c}{Full Network} & \multicolumn{2}{c}{LCC} \\
          & \multicolumn{1}{l}{} & OR    & {95\% CI} & {OR} & {95\% CI} & {OR} & {95\% CI} & {OR} & {95\% CI} & {OR} & {95\% CI} & {OR} & {95\% CI} \\
    \midrule
    \shortstack{Network\\Characteristic} & Edges & 0.01$^\ddagger$ & {(0.008, 0.03)} & {0.05$^\ddagger$} & {(0.03, 0.09)} & {0.01$^\ddagger$} & {(0.009, 0.03)} & {0.05$^\ddagger$} & {(0.02, 0.08)} & {0.02$^\ddagger$} & {(0.009, 0.03)} & {0.05$^\ddagger$} & {(0.03, 0.09)} \\
    \midrule
    \multicolumn{1}{c}{\multirow{3}[2]{*}{Sex}} & Male - Male & \multicolumn{1}{c}{0.83} & {(0.62, 1.12)} & 0.9   & {(0.63, 1.29)} & 0.79  & {(0.59, 1.05)} & 0.86  & {(0.61, 1.21)} & 0.79  & {(0.59, 1.06)} & 0.87  & {(0.62, 1.22)} \\
          & Male - Female & 0.65$^\dagger$ & {(0.48, 0.89)} & {0.58$^\dagger$} & {(0.40, 0.85)} & {0.63$^\dagger$} & {(0.47, 0.85)} & {0.56$^\dagger$} & {(0.40, 0.81)} & {0.63$^\dagger$} & {(0.47, 0.85)} & {0.57$^\dagger$} & {(0.40, 0.81)} \\
          & Female - Female & \multicolumn{12}{c}{Reference Level} \\
    \midrule
    \multicolumn{1}{c}{\multirow{6}[2]{*}{Race}} & African American - African American & 1.43$^\dagger$ & {(1.10, 1.86)} & {1.60$^\dagger$} & {(1.16, 2.21)} & {1.39*} & {(1.07, 1.81)} & {1.57$^\dagger$} & {(1.15, 2.16)} & {1.39*} & {(1.07, 1.80)} & {1.57$^\dagger$} & {(1.14, 2.15)} \\
          & African American - Latinx & 0.23$^\ddagger$ & {(0.16, 0.31)} & {0.49$^\ddagger$} & {(0.34, 0.69)} & {0.22$^\ddagger$} & {(0.16, 0.30)} & {0.46$^\ddagger$} & {(0.32, 0.65)} & {0.22$^\ddagger$} & {(0.16, 0.30)} & {0.46$^\ddagger$} & {(0.32, 0.65)} \\
          & Latinx - Latinx & 0.57$^\ddagger$ & {(0.44, 0.75)} & 1.32  & {(0.94, 1.84)} & {0.57$^\ddagger$} & {(0.44, 0.74)} & 1.21  & {(0.87, 1.68)} & {0.57$^\ddagger$} & {(0.44, 0.74)} & 1.23  & {(0.88, 1.70)} \\
          & African American - White & 0.15$^\ddagger$ & {(0.10, 0.23)} & {0.22$^\ddagger$} & {(0.14, 0.34)} & {0.16$^\ddagger$} & {(0.10, 0.23)} & {0.22$^\ddagger$} & {(0.15, 0.34)} & {0.15$^\ddagger$} & {(0.10, 0.23)} & {0.22$^\ddagger$} & {(0.15, 0.34)} \\
          & Latinx - White & 0.24$^\ddagger$ & {(0.17, 0.32)} & {0.40$^\ddagger$} & {(0.28, 0.57)} & {0.23$^\ddagger$} & {(0.17, 0.31)} & {0.38$^\ddagger$} & {(0.27, 0.54)} & {0.23$^\ddagger$} & {(0.17, 0.31)} & {0.39$^\ddagger$} & {(0.27, 0.55)} \\
          & White - White & \multicolumn{12}{p{45em}}{Reference Level} \\
    \midrule
    \multicolumn{1}{c}{\multirow{3}[2]{*}{Education}} & Less than high school - Less than high school & 1.47$^\dagger$ & {(1.11, 1.96)} & 1.22  & {(0.86, 1.73)} & {1.51$^\dagger$} & {(1.14, 2.01)} & 1.24  & {(0.88, 1.74)} & {1.50$^\dagger$} & {(1.13, 2.00)} & 1.25  & {(0.88, 1.76)} \\
          & Less than high school - High School or more & \multicolumn{1}{c}{0.97} & {(0.73, 1.28)} & 0.91  & {(0.64, 1.28)} & 1.01  & {(0.76, 1.34)} & 0.92  & {(0.65, 1.30)} & 1.01  & {(0.76, 1.34)} & 0.92  & {(0.66, 1.30)} \\
          & High school or more - High school or more & \multicolumn{12}{c}{Reference Level} \\
    \midrule
    \multicolumn{1}{c}{\multirow{3}[2]{*}{Employment}} & Employed - Employed & \multicolumn{1}{c}{1.29} & {(0.58, 2.91)} & 1.37  & {(0.43, 4.33)} & 1.28  & {(0.57, 2.87)} & 1.35  & {(0.43, 4.25)} & 1.27  & {(0.57, 2.86)} & 1.35  & {(0.43, 4.26)} \\
          & Employed  - Unemployed & \multicolumn{1}{c}{0.8} & {(0.62, 1.04)} & 0.89  & {(0.65, 1.21)} & 0.79  & {(0.62, 1.02)} & 0.88  & {(0.65, 1.19)} & 0.79  & {(0.62, 1.02)} & 0.88  & {(0.65, 1.19)} \\
          & Unemployed - Unemployed & \multicolumn{12}{c}{Reference Level} \\
    \midrule
    \multicolumn{1}{c}{\multirow{6}[2]{*}{\shortstack{Living\\Situation}}} & In own place - In own place & 0.35$^\ddagger$ & {(0.23, 0.52)} & {0.41$^\ddagger$} & {(0.25, 0.68)} & {0.34$^\ddagger$} & {(0.23, 0.51)} & {0.42$^\ddagger$} & {(0.26, 0.69)} & {0.35$^\ddagger$} & {(0.23, 0.52)} & {0.42$^\ddagger$} & {(0.26, 0.69)} \\
          & In own place - Someone else's place & 0.27$^\ddagger$ & {(0.19, 0.38)} & {0.30$^\ddagger$} & {(0.20, 0.44)} & {0.28$^\ddagger$} & {(0.20, 0.38)} & {0.31$^\ddagger$} & {(0.21, 0.45)} & {0.26$^\ddagger$} & {(0.19, 0.37)} & {0.29$^\ddagger$} & {(0.20, 0.43)} \\
          & Someone else's place - Someone else's place & 0.33$^\ddagger$ & {(0.24, 0.46)} & {0.36$^\ddagger$} & {(0.25, 0.52)} & {0.33$^\ddagger$} & {(0.24, 0.46)} & {0.37$^\ddagger$} & {(0.26, 0.53)} & {0.33$^\ddagger$} & {(0.23, 0.46)} & {0.35$^\ddagger$} & {(0.25, 0.50)} \\
          & In own place - Homeless & 0.27$^\ddagger$ & {(0.18, 0.41)} & {0.30$^\ddagger$} & {(0.19, 0.47)} & {0.28$^\ddagger$} & {(0.19, 0.41)} & {0.30$^\ddagger$} & {(0.19, 0.47)} & {0.27$^\ddagger$} & {(0.18, 0.40)} & {0.30$^\ddagger$} & {(0.19, 0.47)} \\
          & Someone else's place - Homeless & 0.40$^\ddagger$ & {(0.29, 0.56)} & {0.41$^\ddagger$} & {(0.28, 0.58)} & {0.41$^\ddagger$} & {(0.30, 0.57)} & {0.42$^\ddagger$} & {(0.30, 0.60)} & {0.41$^\ddagger$} & {(0.29, 0.57)} & {0.40$^\ddagger$} & {(0.28, 0.58)} \\
          & Homeless - Homeless & \multicolumn{12}{c}{Reference Level} \\
    \midrule
    \multicolumn{1}{c}{\multirow{6}[2]{*}{\shortstack{Marital\\Status}}} & Single - Single & \multicolumn{1}{c}{1.14} & {(0.81, 1.62)} & {1.58* } & {(1.07, 2.33)} & 1.11  & {(0.78, 1.56)} & {1.62*} & {(1.11, 2.39)} & 1.11  & {(0.78, 1.57)} & {1.64*} & {(1.11, 2.40)} \\
          & Single - Married & \multicolumn{1}{c}{0.8} & {(0.55, 1.17)} & 1.02  & {(0.66, 1.58)} & 0.78  & {(0.53, 1.13)} & 1     & {(0.65, 1.54)} & 0.78  & {(0.54, 1.14)} & 1.03  & {(0.67, 1.58)} \\
          & Married - Married & \multicolumn{1}{c}{1.19} & {(0.72, 1.98)} & 1.55  & {(0.82, 2.91)} & 1.18  & {(0.72, 1.93)} & 1.55  & {(0.85, 2.84)} & 1.2   & {(0.73, 1.96)} & 1.58  & {(0.86, 2.89)} \\
          & Single - Divorced & \multicolumn{1}{c}{0.74} & {(0.51, 1.06)} & 0.83  & {(0.56, 1.24)} & 0.74  & {(0.52, 1.06)} & 0.88  & {(0.60, 1.31)} & 0.75  & {(0.52, 1.06)} & 0.89  & {(0.60, 1.31)} \\
          & Married - Divorced & \multicolumn{1}{c}{0.96} & {(0.63, 1.45)} & 0.91  & {(0.56, 1.49)} & 0.95  & {(0.64, 1.43)} & 0.87  & {(0.53, 1.41)} & 0.97  & {(0.64, 1.45)} & 0.88  & {(0.54, 1.43)} \\
          & Divorced - Divorced & \multicolumn{12}{c}{Reference Level} \\
    \bottomrule
    \end{tabular}%
  \end{center}
  \label{tab:addlabel}%
  \flushleft \footnotesize{Note: node mix models are used to compare pairs of people within and between levels of an attribute variable and compares the pair’s likelihood of having ties compared to a pair reference level.\\
OR = odds ratio\\
95\% CI is the 95\% confidence interval of the odds ratio\\
* indicates $p < 0.05$, $\dagger$ indicates $p < 0.01$, and $\ddagger$ indicates $p < 0.001$}
\end{table}%
\end{sidewaystable}

\begin{sidewaystable}\setlength{\tabcolsep}{1.8pt}
\begin{table}[H]
  \begin{center}\scriptsize
  \caption{ERGM final model results in SFHR for complete cases and both imputation methods, modeled in both the full network and the largest connected component (LCC).}
    \begin{tabular}{clcccccccccccc}
          & \multicolumn{1}{l}{} & \multicolumn{4}{c}{Complete Cases} & \multicolumn{3}{c}{Propensity Score} &       & \multicolumn{4}{c}{Miss Forest} \\
          & \multicolumn{1}{l}{} & \multicolumn{2}{c}{Full Network} & \multicolumn{2}{c}{LCC} & \multicolumn{2}{c}{Full Network} & LCC   &       & \multicolumn{2}{c}{Full Network} & \multicolumn{2}{c}{LCC} \\
          & \multicolumn{1}{l}{} & OR    & 95\% CI & OR    & 95\% CI & OR    & 95\% CI & OR    & 95\% CI & OR    & 95\% CI & OR    & 95\% CI \\
    \midrule
    \shortstack{Network\\Characteristic} & \multicolumn{1}{l}{Edges} & 0.01$^\ddagger$ & (0.007, 0.01) & 0.03$^\ddagger$ & (0.02, 0.04) & 0.01$^\ddagger$ & (0.007, 0.02) & 0.03$^\ddagger$ & (0.02, 0.04) & 0.01$^\ddagger$ & (0.01, 0.02) & 0.03$^\ddagger$ & (0.02, 0.04) \\
    \midrule
    \multirow{6}[2]{*}{Race} & \shortstack{African American -\\African American\ \ \ } & 1.44$^\dagger$ & (1.11, 1.87) & 1.63$^\dagger$ & (1.18, 2.24) & 1.39* & (1.07, 1.80) & 1.59$^\dagger$ & (1.16, 2.19) & 1.40* & (1.08, 1.81) & 1.59$^\dagger$ & (1.16, 2.18) \\
          & \multicolumn{1}{l}{African American - Latinx} & 0.23$^\ddagger$ & (0.17, 0.31) & 0.49$^\ddagger$ & (0.34, 0.70) & 0.22$^\ddagger$ & (0.16, 0.30) & 0.46$^\ddagger$ & (0.33, 0.66) & 0.22$^\ddagger$ & (0.16, 0.30) & 0.46$^\ddagger$ & (0.33, 0.66) \\
          & \multicolumn{1}{l}{Latinx - Latinx} & 0.58$^\ddagger$ & (0.44, 0.76) & 1.33  & (0.95, 1.86) & 0.58$^\ddagger$ & (0.45, 0.75) & 1.26  & (0.91, 1.74) & 0.58$^\ddagger$ & (0.45, 0.75) & 1.25  & (0.90, 1.73) \\
          & \multicolumn{1}{l}{African American - White} & 0.15$^\ddagger$ & (0.10, 0.23) & 0.22$^\ddagger$ & (0.14, 0.34) & 0.15$^\ddagger$ & (0.10, 0.23) & 0.23$^\ddagger$ & (0.15, 0.35) & 0.16$^\ddagger$ & (0.10, 0.23) & 0.23$^\ddagger$ & (0.15, 0.35) \\
          & \multicolumn{1}{l}{Latinx - White} & 0.24$^\ddagger$ & (0.18, 0.32) & 0.40$^\ddagger$ & (0.28, 0.57) & 0.24$^\ddagger$ & (0.18, 0.32) & 0.39$^\ddagger$ & (0.27, 0.55) & 0.24$^\ddagger$ & (0.18, 0.32) & 0.39$^\ddagger$ & (0.27, 0.55) \\
          & \multicolumn{1}{l}{White - White} & \multicolumn{12}{c}{Reference Level} \\
    \midrule
    \multirow{6}[2]{*}{\shortstack{Living\\Situation}} & \multicolumn{1}{l}{Own place - Own place} & 0.35$^\ddagger$ & (0.23, 0.52) & 0.42$^\ddagger$ & (0.26, 0.69) & 0.34$^\ddagger$ & (0.23, 0.50) & 0.41$^\ddagger$ & (0.25, 0.66) & 0.35$^\ddagger$ & (0.23, 0.52) & 0.43$^\ddagger$ & (0.26, 0.69) \\
          & \multicolumn{1}{l}{Own place - Someone else} & 0.27$^\ddagger$ & (0.19, 0.37) & 0.30$^\ddagger$ & (0.21, 0.44) & 0.27$^\ddagger$ & (0.19, 0.37) & 0.31$^\ddagger$ & (0.22, 0.45) & 0.26$^\ddagger$ & (0.19, 0.36) & 0.30$^\ddagger$ & (0.20, 0.43) \\
          & \multicolumn{1}{l}{Someone else - Someone else} & 0.33$^\ddagger$ & (0.23, 0.46) & 0.36$^\ddagger$ & (0.25, 0.52) & 0.33$^\ddagger$ & (0.24, 0.46) & 0.37$^\ddagger$ & (0.26, 0.53) & 0.32$^\ddagger$ & (0.23, 0.45) & 0.35$^\ddagger$ & (0.25, 0.50) \\
          & \multicolumn{1}{l}{Own place - Homeless} & 0.27$^\ddagger$ & (0.18, 0.41) & 0.30$^\ddagger$ & (0.19, 0.48) & 0.28$^\ddagger$ & (0.19, 0.41) & 0.30$^\ddagger$ & (0.19, 0.46) & 0.27$^\ddagger$ & (0.18, 0.41) & 0.30$^\ddagger$ & (0.19, 0.47) \\
          & \multicolumn{1}{l}{Someone else - Homeless} & 0.40$^\ddagger$ & (0.28, 0.56) & 0.41$^\ddagger$ & (0.28, 0.58) & 0.40$^\ddagger$ & (0.29, 0.56) & 0.41$^\ddagger$ & (0.29, 0.59) & 0.40$^\ddagger$ & (0.29, 0.56) & 0.40$^\ddagger$ & (0.28, 0.58) \\
          & \multicolumn{1}{l}{Homeless - Homeless} & \multicolumn{12}{c}{Reference Level} \\
    \midrule
    \multirow{2}[2]{*}{\shortstack{Sex\\homophily}} & \multicolumn{1}{l}{Male} & 1.27* & (1.05, 1.54) & 1.52$^\ddagger$ & (1.21, 1.92) & 1.24* & (1.02, 1.49) & 1.51$^\ddagger$ & (1.21, 1.90) & 1.24* & (1.03, 1.50) & 1.51$^\ddagger$ & (1.21, 1.90) \\
          & \multicolumn{1}{l}{Female} & 1.55$^\dagger$ & (1.13, 2.11) & 1.73$^\dagger$ & (1.19, 2.52) & 1.59$^\dagger$ & (1.18, 2.14) & 1.76$^\dagger$ & (1.23, 2.52) & 1.59$^\dagger$ & (1.18, 2.14) & 1.76$^\dagger$ & (1.23, 2.52) \\
    \midrule
    \multirow{3}[2]{*}{\shortstack{Marital\\status\\homophily}} & \multicolumn{1}{l}{Single} & 1.44$^\ddagger$ & (1.19, 1.76) & 1.77$^\ddagger$ & (1.40, 2.23) & 1.40$^\ddagger$ & (1.15, 1.70) & 1.75$^\ddagger$ & (1.39, 2.20) & 1.40$^\ddagger$ & (1.15, 1.70) & 1.77$^\ddagger$ & (1.41, 2.23) \\
          & \multicolumn{1}{l}{In relationship} & 1.48  & (0.98, 2.23) & 1.7   & (0.99, 2.94) & 1.50* & (1.01, 2.22) & 1.71* & (1.02, 2.87) & 1.49* & (1.002, 2.21) & 1.69* & (1.01, 2.83) \\
          & \multicolumn{1}{l}{Divorced} & 1.26  & (0.90, 1.76) & 1.12  & (0.77, 1.62) & 1.26  & (0.90, 1.76) & 1.09  & (0.76, 1.58) & 1.25  & (0.90, 1.75) & 1.08  & (0.75, 1.56) \\
    \midrule
    \multicolumn{1}{c}{\multirow{2}[2]{*}{Education}} & High School or more & 0.77$^\ddagger$ & (0.67, 0.88) & 0.85* & (0.72, 0.998) & 0.77$^\ddagger$ & (0.67, 0.88) & 0.84* & (0.71, 0.99) & 0.77$^\ddagger$ & (0.67, 0.88) & {0.84*} & {(0.71, 0.99)} \\
          & Less than High School & \multicolumn{12}{c}{Reference Level} \\
    \bottomrule
    \end{tabular}%
    \end{center}
  \label{tab:addlabel}%
  \flushleft \footnotesize{OR = odds ratio\\
95\% CI is the 95\% confidence interval of the odds ratio\\
* indicates $p < 0.05$, $\dagger$ indicates $p < 0.01$, and $\ddagger$ indicates $p < 0.001$}
\end{table}%
\end{sidewaystable}

\begin{sidewaystable}
\begin{table}[H]
  \centering\small
  \caption{ERGM node match model and node factor model with geometrically-weighted degree results in SFHR largest connected component using complete case analysis}
    \begin{tabular}{clcccccccc}
          & \multicolumn{1}{l}{} & \multicolumn{4}{c}{Node match model} & \multicolumn{4}{c}{Node factor model} \\
          & \multicolumn{1}{r}{} & \multicolumn{2}{c}{With gwdegree} & \multicolumn{2}{c}{Without gwdegree} & \multicolumn{2}{c}{With gwdegree} & \multicolumn{2}{c}{Without gwdegree} \\
          & \multicolumn{1}{r}{} & OR    & 95\% CI & OR    & 95\% CI & OR    & 95\% CI & OR    & 95\% CI \\
    \midrule
    {\multirow{2}[2]{*}{\shortstack{Network\\Characteristic}}} & Edges & 0.005$^\ddagger$ & (0.003, 0.008) & 0.002$^\ddagger$ & (0.0016, 0.003) & 0.02$^\ddagger$ & (0.01, 0.04) & {0.01$^\ddagger$} & {(0.007, 0.02)} \\
          & \shortstack{Geometrically-\\weighted degree} & 0.41$^\ddagger$ & (0.28, 0.59) & {-} & {-} & 0.43$^\ddagger$ & (0.30, 0.62) & -     & - \\
    \midrule
    {\multirow{2}[2]{*}{Sex}} & Female & 1.81$^\dagger$ & (1.26, 2.62) & 1.72$^\dagger$ & (1.18, 2.51) & 0.9   & (0.78, 1.04) & 0.87  & {(0.73, 1.03)} \\
          & Male  & 1.48$^\ddagger$ & (1.20, 1.83) & 1.54$^\ddagger$ & (1.22, 1.95) & \multicolumn{4}{c}{Reference Level} \\
    \midrule
    {\multirow{3}[2]{*}{Race}} & Latinx & 3.65$^\ddagger$ & (2.80, 4.77) & 3.65$^\ddagger$ & (2.73, 4.88) & 0.99  & (0.85, 1.15) & 0.99  & {(0.83, 1.18)} \\
          & White & 2.99$^\ddagger$ & (2.28, 3.91) & 2.77$^\ddagger$ & (2.07, 3.70) & 0.83* & (0.71, 0.97) & {0.77$^\dagger$} & {(0.64, 0.92)} \\
          & African American & 4.17$^\ddagger$ & (3.24, 5.36) & 4.44$^\ddagger$ & (3.36, 5.85) & \multicolumn{4}{c}{Reference Level} \\
    \midrule
    {\multirow{2}[2]{*}{Education}} & High School or more & {1.13} & (0.82, 1.57) & {1.07} & (0.76, 1.52) & 0.89  & (0.77, 1.02) & {0.85* } & {(0.72, 0.996)} \\
          & Less than High School & 1.31* & (1.06, 1.61) & 1.39$^\dagger$ & (1.11, 1.74) & \multicolumn{4}{c}{Reference Level} \\
    \midrule
    {\multirow{2}[2]{*}{Employment}} & Unemployed & {1.11} & (0.85, 1.45) & {1.15} & (0.85, 1.56) & 1.05  & (0.83, 1.34) & 1.07  & {(0.81, 1.42)} \\
          & Employed & {1.56} & (0.50, 4.85) & {1.5} & (0.46, 4.88) & \multicolumn{4}{c}{Reference Level} \\
    \midrule
    {\multirow{3}[2]{*}{\shortstack{Living\\Situation}}} & Someone else's place & {1.12} & (0.88, 1.42) & {1.04} & (0.81, 1.35) & 1.06  & (0.90, 1.25) & 1.08  & {(0.88, 1.31)} \\
          & Homeless & 2.54$^\ddagger$ & (1.86, 3.46) & 2.89$^\ddagger$ & (2.08, 4.01) & 1.37$^\ddagger$ & (1.14, 1.65) & {1.54$^\ddagger$} & {(1.24, 1.91)} \\
          & Own place & {1.31} & (0.89, 1.93) & {1.24} & (0.82, 1.87) & \multicolumn{4}{c}{Reference Level} \\
    \midrule
    {\multirow{3}[2]{*}{\shortstack{Marital\\Status}}} & In relationship & 1.72* & (1.03, 2.86) & {1.63} & (0.95, 2.81) & 0.9   & (0.75, 1.08) & 0.87  & {(0.70, 1.08)} \\
          & Divorced & {1.21} & (0.85, 1.72) & {1.11} & (0.77, 1.62) & 0.79$^\dagger$ & (0.68, 0.92) & {0.72$^\ddagger$} & {(0.61, 0.86)} \\
          & Single & 1.65$^\ddagger$ & (1.33, 2.05) & 1.83$^\ddagger$ & (1.45, 2.31) & \multicolumn{4}{c}{Reference Level} \\
    \bottomrule
    \end{tabular}%
  \label{tab:addlabel}%
  \flushleft \footnotesize{Note: the node match model represents the odds of two people being connected in a network based on sharing some attributes, e.g., males being connected to other males in the network. The node factor model is used to compare people across levels of an attribute variable to see if people are more or less likely to have ties in the network, compared to a reference level.\\
OR = odds ratio\\
95\% CI is the 95\% confidence interval of the odds ratio\\
* indicates $p < 0.05$, $\dagger$ indicates $p < 0.01$, and $\ddagger$ indicates $p < 0.001$}
\end{table}%
\end{sidewaystable}

\begin{sidewaystable}\setlength{\tabcolsep}{1.8pt}
\begin{table}[H]
  \begin{center}\scriptsize
  \caption{ERGM node match model results in TRIP for complete cases and both imputation methods, modeled in both the full network and the largest connected component (LCC).}
    \begin{tabular}{clcrcrcrcrcrcr}
          & \multicolumn{1}{r}{} & \multicolumn{4}{c}{Complete Cases} & \multicolumn{4}{c}{Propensity Score Matching} & \multicolumn{4}{c}{Miss Forest} \\
          & \multicolumn{1}{r}{} & \multicolumn{2}{c}{Full Network} & \multicolumn{2}{c}{LCC} & \multicolumn{2}{c}{Full Network} & \multicolumn{2}{c}{LCC} & \multicolumn{2}{c}{Full Network} & \multicolumn{2}{c}{LCC} \\
          & \multicolumn{1}{r}{} & OR    & 95\% CI & OR    & 95\% CI & OR    & 95\% CI & OR    & 95\% CI & OR    & 95\% CI & OR    & 95\% CI \\
    \midrule
    \shortstack{Network\\Characteristic} & Edges & 0.006$^\ddagger$ & (0.004, 0.008) & 0.011$^\ddagger$ & (0.009, 0.015) & 0.006$^\ddagger$ & (0.004, 0.008) & 0.01$^\ddagger$ & (0.008, 0.015) & 0.006$^\ddagger$ & (0.004, 0.008) & 0.01$^\ddagger$ & (0.008, 0.01) \\
    \midrule
    {\multirow{2}[2]{*}{Sex}} & Male  & {1.01} & (0.84, 1.23) & {1.04} & (0.85, 1.27) & {0.99} & (0.82, 1.19) & {0.99} & (0.81, 1.20) & {0.99} & (0.82, 1.19) & {0.99} & (0.81, 1.20) \\
          & Female & 1.73$^\dagger$ & (1.22, 2.47) & 1.94$^\ddagger$ & (1.35, 2.79) & 1.92$^\ddagger$ & (1.38, 2.67) & 2.15$^\ddagger$ & (1.53, 3.03) & 1.92$^\ddagger$ & (1.38, 2.67) & 2.16$^\ddagger$ & (1.53, 3.04) \\
    \midrule
    {\multirow{2}[2]{*}{Nationality}} & Greek & {1.19} & (0.94, 1.52) & 1.42$^\dagger$ & (1.11, 1.83) & {1.18} & (0.93, 1.50) & 1.43$^\dagger$ & (1.12, 1.83) & {1.2} & (0.95, 1.53) & 1.45$^\dagger$ & (1.14, 1.86) \\
          & Not Greek & 2.88$^\ddagger$ & (1.71, 4.85) & 2.43$^\dagger$ & (1.43, 4.15) & 3.04$^\ddagger$ & (1.83, 5.06) & 2.49$^\ddagger$ & (1.48, 4.18) & 2.89$^\ddagger$ & (1.74, 4.80) & 2.38$^\dagger$ & (1.41, 4.00) \\
    \midrule
    {\multirow{3}[2]{*}{Education}} & Primary School & {1.08} & (0.82, 1.42) & {0.91} & (0.67, 1.24) & {1.12} & (0.86, 1.46) & {0.96} & (0.71, 1.29) & {1.12} & (0.86, 1.45) & {0.96} & (0.71, 1.29) \\
          & High School & {0.86} & (0.70, 1.06) & {0.82} & (0.66, 1.02) & {0.89} & (0.73, 1.09) & {0.83} & (0.67, 1.03) & {0.89} & (0.73, 1.09) & {0.83} & (0.68, 1.03) \\
          & Post High School & {1.6} & (0.95, 2.71) & {1.43} & (0.81, 2.52) & {1.6} & (0.95, 2.71) & {1.5} & (0.85, 2.63) & {1.6} & (0.94, 2.70) & {1.49} & (0.85, 2.63) \\
    \midrule
    {\multirow{4}[2]{*}{Employment}} & Employed & {0.53} & (0.24, 1.19) & {0.64} & (0.29, 1.46) & {0.49} & (0.22, 1.10) & {0.59} & (0.26, 1.34) & {0.5} & (0.22, 1.12) & {0.6} & (0.26, 1.34) \\
          & \shortstack{Unemployed:\ \ \ \ \ \  \\looking for work} & {1.12} & (0.76, 1.65) & {1.23} & (0.79, 1.91) & {1.16} & (0.80, 1.67) & {1.23} & (0.80, 1.89) & {1.15} & (0.80, 1.66) & {1.22} & (0.79, 1.88) \\
          & \shortstack{Can't work,\ \ \ \ \\health reasons} & 1.28*  & (1.05, 1.57) & {1.18} & (0.96, 1.45) & 1.28* & (1.04, 1.56) & {1.18} & (0.96, 1.45) & 1.26* & (1.03, 1.54) & {1.16} & (0.95, 1.43) \\
          & Other & 1.67*  & (1.01, 2.74) & {1.6} & (0.96, 2.66) & {1.57} & (0.96, 2.59) & {1.43} & (0.86, 2.38) & {1.59} & (0.97, 2.60) & {1.43} & (0.86, 2.38) \\
    \midrule
    {\multirow{3}[2]{*}{\shortstack{Living\\Situation}}} & Paying rent & {0.82} & (0.49, 1.36) & {1.04} & (0.58, 1.88) & {0.83} & (0.51, 1.37) & {1.02} & (0.58, 1.81) & {0.79} & (0.48, 1.32) & {0.96} & (0.54, 1.74) \\
          & Not paying rent & {1.03} & (0.84, 1.27) & {0.98} & (0.78, 1.23) & {1.03} & (0.83, 1.26) & {0.99} & (0.80, 1.24) & {1.05} & (0.85, 1.28) & {1.01} & (0.81, 1.26) \\
          & Homeless & 3.88$^\ddagger$ & (3.02, 4.98) & 2.63$^\ddagger$ & (2.04, 3.40) & 3.77$^\ddagger$ & (2.95, 4.83) & 2.53$^\ddagger$ & (1.97, 3.25) & 3.84$^\ddagger$ & (3.00, 4.92) & 2.58$^\ddagger$ & (2.01, 3.33) \\
    \bottomrule
    \end{tabular}%
  \end{center}
  \label{tab:addlabel}%
  \flushleft \footnotesize{Note: the node match model represents the odds of two people being connected in a network based on sharing some attributes, e.g., males being connected to other males in the network.\\
OR = odds ratio\\
95\% CI is the 95\% confidence interval of the odds ratio\\
* indicates $p < 0.05$, $\dagger$ indicates $p < 0.01$, and $\ddagger$ indicates $p < 0.001$}
\end{table}%
\end{sidewaystable}

\begin{sidewaystable}\setlength{\tabcolsep}{2pt}
\begin{table}[H]
  \begin{center}\scriptsize
  \caption{ERGM node factor model results in TRIP for complete cases and both imputation methods, modeled in both the full network and the largest connected component (LCC).}
    \begin{tabular}{clcrcrcrcrcrcr}
          & \multicolumn{1}{r}{} & \multicolumn{4}{c}{Complete Cases} & \multicolumn{4}{c}{Propensity Score Matching} & \multicolumn{4}{c}{Miss Forest} \\
          & \multicolumn{1}{r}{} & \multicolumn{2}{c}{Full Network} & \multicolumn{2}{c}{LCC} & \multicolumn{2}{c}{Full Network} & \multicolumn{2}{c}{LCC} & \multicolumn{2}{c}{Full Network} & \multicolumn{2}{c}{LCC} \\
          & \multicolumn{1}{r}{} & OR    & {95\% CI} & {OR} & {95\% CI} & {OR} & {95\% CI} & {OR} & {95\% CI} & {OR} & {95\% CI} & {OR} & {95\% CI} \\
    \midrule
    \shortstack{Network\\Characteristic} & {Edges} & 0.002$^\ddagger$ & {(0.001, 0.004)} & {0.006$^\ddagger$} & {(0.004, 0.01)} & {{0.002$^\ddagger$}} & {{(0.001, 0.003)}} & {{0.005$^\ddagger$}} & {{(0.003, 0.009)}} & {{0.002$^\ddagger$}} & {{(0.001, 0.003)}} & {{0.005$^\ddagger$}} & {{(0.003, 0.009)}} \\
    \midrule
    \multicolumn{1}{c}{\multirow{2}[2]{*}{{Sex}}} & {Female} & \multicolumn{1}{c}{1.12} & {(0.96, 1.30)} & {1.18* } & {(1.01, 1.38)} & {{1.18*}} & {{(1.02, 1.36)}} & {{1.27$^\dagger$}} & {{(1.09, 1.48)}} & {{1.18*}} & {{(1.02, 1.36)}} & {{1.27$^\dagger$}} & {{(1.09, 1.48)}} \\
          & {Male} & \multicolumn{12}{c}{Reference Level} \\
    \midrule
    \multicolumn{1}{c}{\multirow{2}[2]{*}{{Nationality}}} & {Not Greek} & \multicolumn{1}{c}{0.996} & {(0.81, 1.22)} & 0.88  & {(0.71, 1.09)} & {1.02} & {{(0.83, 1.24)}} & {0.88} & {{(0.72, 1.09)}} & {0.99} & {{(0.81, 1.21)}} & {0.87} & {{(0.71, 1.07)}} \\
          & {Greek} & \multicolumn{12}{c}{Reference Level} \\
    \midrule
    \multicolumn{1}{c}{\multirow{3}[2]{*}{{Education}}} & High School & \multicolumn{1}{c}{0.97} & {(0.84, 1.12)} & 0.99  & {(0.85, 1.15)} & {0.98} & {{(0.85, 1.12)}} & {0.99} & {{(0.85, 1.15)}} & {0.98} & {{(0.85, 1.12)}} & {0.99} & {{(0.86, 1.15)}} \\
          & \shortstack{Post High\\School\ \ \ \ \ \ } & 1.35$^\dagger$  & {(1.12, 1.63)} & {1.36$^\dagger$} & {(1.12, 1.65)} & {{1.34$^\dagger$}} & {{(1.12, 1.61)}} & {{1.39$^\ddagger$}} & {{(1.14, 1.68)}} & {{1.34$^\dagger$}} & {{(1.11, 1.61)}} & {{1.38$^\ddagger$}} & {{(1.14, 1.68)}} \\
          & \shortstack{Less than\ \ \ \ \\high school} & \multicolumn{12}{c}{Reference Level} \\
    \midrule
    \multicolumn{1}{c}{\multirow{4}[2]{*}{{Employment}}} & \shortstack{Unemployed:\ \ \ \ \ \ \\looking for work} & \multicolumn{1}{c}{1.25} & {(0.99, 1.57)} & {1.40$^\dagger$} & {(1.10, 1.78)} & {{1.36$^\dagger$}} & {{(1.09, 1.70)}} & {{1.51$^\ddagger$}} & {{(1.19, 1.92)}} & {{1.32*}} & {{(1.05, 1.65)}} & {{1.48$^\dagger$}} & {{(1.17, 1.87)}} \\
          & \shortstack{Can't work,\ \ \ \\health reasons} & 1.34$^\dagger$ & {(1.08, 1.66)} & {1.31* } & {(1.04, 1.63)} & {{1.39$^\dagger$}} & {{(1.13, 1.72)}} & {{1.35$^\dagger$}} & {{(1.08, 1.67)}} & {{1.35$^\dagger$}} & {{(1.09, 1.67)}} & {{1.31*}} & {{(1.05, 1.64)}} \\
          & Other & 1.32*  & {(1.02, 1.71)} & {1.31* } & {(1.01, 1.71)} & {{1.39$^\dagger$}} & {{(1.09, 1.79)}} & {{1.35*}} & {{(1.04, 1.74)}} & {{1.36*}} & {{(1.06, 1.74)}} & {{1.32*}} & {{(1.02, 1.71)}} \\
          & Employed & \multicolumn{12}{c}{Reference Level} \\
    \midrule
    \multicolumn{1}{c}{\multirow{3}[2]{*}{\shortstack{Living\\Situation}}} & Not paying rent & 1.29$^\dagger$ & {(1.06, 1.56)} & 1.08  & {(0.88, 1.33)} & {{1.25*}} & {{(1.04, 1.51)}} & {1.08} & {{(0.88, 1.32)}} & {{1.31*}} & {{(1.09, 1.58)}} & {1.12} & {{(0.92, 1.38)}} \\
          & Homeless & 2.35$^\ddagger$ & {(1.91, 2.89)} & {1.72$^\ddagger$} & {(1.38, 2.14)} & {{2.27$^\ddagger$}} & {{(1.85, 2.77)}} & {{1.70$^\ddagger$}} & {{(1.37, 2.10)}} & {{2.37$^\ddagger$}} & {{(1.93, 2.91)}} & {{1.77$^\ddagger$}} & {{(1.42, 2.19)}} \\
          & Paying rent & \multicolumn{12}{c}{Reference Level} \\
    \bottomrule
    \end{tabular}%
    \end{center}
  \label{tab:addlabel}%
  \flushleft \footnotesize{Note: the node factor model is used to compare people across levels of an attribute variable to see if people are more or less likely to have ties in the network, compared to a reference level.\\
OR = odds ratio\\
95\% CI is the 95\% confidence interval of the odds ratio\\
* indicates $p < 0.05$, $\dagger$ indicates $p < 0.01$, and $\ddagger$ indicates $p < 0.001$}
\end{table}%
\end{sidewaystable}

\begin{sidewaystable}\setlength{\tabcolsep}{2.5pt}
\begin{table}[H]
  \begin{center}\scriptsize
  \caption{ERGM node mix model results in TRIP for complete cases and both imputation methods, modeled in both the full network and the largest connected component (LCC).}
    \begin{tabular}{clcccccccccccc}
          & \multicolumn{1}{r}{} & \multicolumn{4}{c}{Complete Cases} & \multicolumn{4}{c}{Propensity Score Matching} & \multicolumn{4}{c}{Miss Forest} \\
          & \multicolumn{1}{r}{} & \multicolumn{2}{c}{Full Network} & \multicolumn{2}{c}{LCC} & \multicolumn{2}{c}{Full Network} & \multicolumn{2}{c}{LCC} & \multicolumn{2}{c}{Full Network} & \multicolumn{2}{c}{LCC} \\
          & \multicolumn{1}{r}{} & OR    & {95\% CI} & {OR} & {95\% CI} & {OR} & {95\% CI} & {OR} & {95\% CI} & {OR} & {95\% CI} & {OR} & {95\% CI} \\
    \midrule
    \shortstack{Network\\Characteristic} & Edges & 0.27$^\dagger$ & {(0.12, 0.63)} & {0.33* } & {(0.14, 0.82)} & {0.30$^\dagger$} & {(0.13, 0.68)} & {0.33*} & {(0.14, 0.82)} & {0.28$^\dagger$} & {(0.12, 0.65)} & {0.33*} & {(0.13, 0.79)} \\
    \midrule
    \multicolumn{1}{c}{\multirow{3}[2]{*}{Sex}} & Male - Male & 0.61$^\dagger$ & {(0.43, 0.86)} & {0.51$^\ddagger$} & {(0.36, 0.74)} & {0.53$^\ddagger$} & {(0.39, 0.74)} & {0.44$^\ddagger$} & {(0.31, 0.62)} & {0.53$^\ddagger$} & {(0.39, 0.74)} & {0.44$^\ddagger$} & {(0.32, 0.61)} \\
          & Male - Female & 0.59$^\dagger$ & {(0.41, 0.83)} & {0.51$^\ddagger$} & {(0.35, 0.73)} & {0.53$^\ddagger$} & {(0.38, 0.74)} & {0.45$^\ddagger$} & {(0.32, 0.64)} & {0.53$^\ddagger$} & {(0.38, 0.74)} & {0.46$^\ddagger$} & {(0.32, 0.64)} \\
          & Female - Female & \multicolumn{12}{c}{Reference Level} \\
    \midrule
    \multicolumn{1}{c}{\multirow{3}[2]{*}{Nationality}} & Greek - Greek & 0.44$^\dagger$ & {(0.27, 0.74)} & {0.58* } & {(0.34, 0.98)} & {0.41$^\ddagger$} & {(0.25, 0.68)} & {0.56*} & {(0.34, 0.94)} & {0.44$^\dagger$} & {(0.27, 0.72)} & {0.60*} & {(0.36, 0.99)} \\
          & Greek - Not Greek & 0.34$^\ddagger$ & {(0.20, 0.58)} & {0.40$^\ddagger$} & {(0.23, 0.69)} & {0.32$^\ddagger$} & {(0.20, 0.54)} & {0.39$^\ddagger$} & {(0.23, 0.66)} & {0.34$^\ddagger$} & {(0.20, 0.56)} & {0.40$^\ddagger$} & {(0.24, 0.68)} \\
          & Not Greek - Not Greek & \multicolumn{12}{c}{Reference Level} \\
    \midrule
    \multicolumn{1}{c}{\multirow{6}[2]{*}{Education}} & Primary School - Primary School & \multicolumn{1}{c}{0.63} & {(0.36, 1.13)} & 0.6   & {(0.32, 1.13)} & 0.66  & {(0.37, 1.16)} & 0.6   & {(0.32, 1.12)} & 0.66  & {(0.37, 1.16)} & 0.6   & {(0.32, 1.12)} \\
          & Primary School - High School & 0.52*  & {(0.30, 0.89)} & 0.59  & {(0.33, 1.05)} & {0.52*} & {(0.30, 0.89)} & {0.55*} & {(0.31, 0.99)} & {0.52*} & {(0.30, 0.89)} & {0.56*} & {(0.31, 0.99)} \\
          & High School - High School & 0.55*  & {(0.32, 0.94)} & 0.59  & {(0.33, 1.05)} & {0.57*} & {(0.33, 0.98)} & 0.57  & {(0.31, 1.01)} & {0.57*} & {(0.33, 0.98)} & 0.57  & {(0.32, 1.02)} \\
          & Primary School - Post High School & \multicolumn{1}{c}{0.73} & {(0.41, 1.31)} & 0.83  & {(0.45, 1.53)} & 0.72  & {(0.41, 1.28)} & 0.8   & {(0.43, 1.48)} & 0.72  & {(0.41, 1.29)} & 0.8   & {(0.43, 1.48)} \\
          & High School - Post High School & \multicolumn{1}{c}{0.77} & {(0.44, 1.34)} & 0.83  & {(0.46, 1.50)} & 0.79  & {(0.45, 1.37)} & 0.81  & {(0.45, 1.47)} & 0.79  & {(0.46, 1.37)} & 0.82  & {(0.45, 1.47)} \\
          & Post High School - Post High School & \multicolumn{12}{c}{Reference Level} \\
    \midrule
    \multicolumn{1}{c}{\multirow{10}[2]{*}{Employment}} & Employed - Employed & 0.38*  & {(0.15, 0.98)} & 0.44  & {(0.17, 1.16)} & {0.37*} & {(0.14, 0.94)} & 0.46  & {(0.18, 1.19)} & {0.38*} & {(0.15, 0.97)} & 0.47  & {(0.18, 1.22)} \\
          & Employed - Unemployed & 0.51*  & {(0.27, 0.96)} & 0.58  & {(0.30, 1.12)} & {0.53*} & {(0.29, 1.00)} & 0.66  & {(0.34, 1.27)} & 0.54  & {(0.29, 1.01)} & 0.67  & {(0.35, 1.29)} \\
          & Unemployed - Unemployed & {0.76} & {(0.41, 1.40)} & 0.84  & {(0.43, 1.64)} & 0.83  & {(0.45, 1.52)} & 0.95  & {(0.49, 1.85)} & 0.82  & {(0.45, 1.50)} & 0.95  & {(0.49, 1.83)} \\
          & Employed - Can't work & {0.58} & {(0.33, 1.01)} & {0.54* } & {(0.30, 0.96)} & {0.57*} & {(0.33, 1.00)} & 0.57  & {(0.32, 1.01)} & 0.59  & {(0.34, 1.02)} & 0.58  & {(0.33, 1.03)} \\
          & Unemployed - Can't work & {0.7} & {(0.42, 1.19)} & 0.76  & {(0.44, 1.31)} & 0.76  & {(0.45, 1.27)} & 0.88  & {(0.51, 1.50)} & 0.75  & {(0.45, 1.26)} & 0.87  & {(0.51, 1.49)} \\
          & Can't work - Can't work & {0.83} & {(0.50, 1.38)} & 0.77  & {(0.46, 1.31)} & 0.87  & {(0.52, 1.45)} & 0.86  & {(0.51, 1.46)} & 0.86  & {(0.52, 1.43)} & 0.85  & {(0.51, 1.44)} \\
          & Employed - Other & {0.72} & {(0.38, 1.37)} & 0.64  & {(0.33, 1.28)} & 0.72  & {(0.38, 1.37)} & 0.68  & {(0.35, 1.34)} & 0.74  & {(0.39, 1.41)} & 0.7   & {(0.36, 1.37)} \\
          & Unemployed - Other & {0.69} & {(0.38, 1.25)} & 0.83  & {(0.45, 1.53)} & 0.83  & {(0.47, 1.47)} & 1.06  & {(0.59, 1.91)} & 0.82  & {(0.46, 1.46)} & 1.05  & {(0.58, 1.90)} \\
          & Can?t work - Other & {0.66} & {(0.39, 1.12)} & 0.61  & {(0.35, 1.06)} & 0.68  & {(0.40, 1.16)} & 0.66  & {(0.38, 1.13)} & 0.68  & {(0.40, 1.14)} & 0.65  & {(0.38, 1.12)} \\
          & Other - Other & \multicolumn{12}{c}{Reference Level} \\
    \midrule
    {\multirow{6}[2]{*}{\shortstack{Living\\Situation}}} & Paying rent - Paying rent & 0.21$^\ddagger$ & {(0.12, 0.36)} & {0.39$^\dagger$} & {(0.21, 0.73)} & {0.22$^\ddagger$} & {(0.13, 0.38)} & {0.39$^\dagger$} & {(0.21, 0.71)} & {0.20$^\ddagger$} & {(0.12, 0.36)} & {0.36$^\dagger$} & {(0.19, 0.68)} \\
          & Paying rent - Not paying rent & 0.18$^\ddagger$ & {(0.13, 0.25)} & {0.30$^\ddagger$} & {(0.21, 0.42)} & {0.18$^\ddagger$} & {(0.13, 0.25)} & {0.30$^\ddagger$} & {(0.21, 0.43)} & {0.18$^\ddagger$} & {(0.13, 0.25)} & {0.30$^\ddagger$} & {(0.21, 0.42)} \\
          & Not paying rent - Not paying rent & 0.25$^\ddagger$ & {(0.19, 0.34)} & {0.35$^\ddagger$} & {(0.25, 0.48)} & {0.26$^\ddagger$} & {(0.19, 0.34)} & {0.36$^\ddagger$} & {(0.26, 0.49)} & {0.25$^\ddagger$} & {(0.19, 0.34)} & {0.36$^\ddagger$} & {(0.26, 0.49)} \\
          & Paying rent - Homeless & 0.23$^\ddagger$ & {(0.16, 0.34)} & {0.37$^\ddagger$} & {(0.25, 0.54)} & {0.25$^\ddagger$} & {(0.17, 0.36)} & {0.38$^\ddagger$} & {(0.26, 0.55)} & {0.23$^\ddagger$} & {(0.16, 0.34)} & {0.35$^\ddagger$} & {(0.24, 0.52)} \\
          & Not paying rent - Homeless & 0.32$^\ddagger$ & {(0.25, 0.42)} & {0.41$^\ddagger$} & {(0.31, 0.54)} & {0.32$^\ddagger$} & {(0.25, 0.43)} & {0.42$^\ddagger$} & {(0.32, 0.55)} & {0.33$^\ddagger$} & {(0.25, 0.43)} & {0.42$^\ddagger$} & {(0.32, 0.55)} \\
          & Homeless - Homeless & \multicolumn{12}{c}{Reference Level} \\
    \bottomrule
    \end{tabular}%
    \end{center}
  \label{tab:addlabel}%
  \flushleft \footnotesize{Note: node mix models are used to compare pairs of people within and between levels of an attribute variable and compares the pair’s likelihood of having ties compared to a pair reference level.\\
OR = odds ratio\\
95\% CI is the 95\% confidence interval of the odds ratio\\
* indicates $p < 0.05$, $\dagger$ indicates $p < 0.01$, and $\ddagger$ indicates $p < 0.001$}
\end{table}%
\end{sidewaystable}

\begin{sidewaystable}\setlength{\tabcolsep}{2.5pt}
\begin{table}[H]
  \begin{center}\scriptsize
  \caption{ERGM final model results in TRIP for complete cases and both imputation methods, modeled in both the full network and the largest connected component (LCC).}
    \begin{tabular}{clcccccccccccc}
          &       & \multicolumn{4}{c}{Complete Cases} & \multicolumn{4}{c}{Propensity Score} & \multicolumn{4}{c}{Miss Forest} \\
          &       & \multicolumn{2}{c}{Full Network} & \multicolumn{2}{c}{LCC} & \multicolumn{2}{c}{Full Network} & LCC   &       & \multicolumn{2}{c}{Full Network} & \multicolumn{2}{c}{LCC} \\
          &       & OR    & 95\% CI & OR    & 95\% CI & OR    & 95\% CI & OR    & 95\% CI & OR    & 95\% CI & OR    & 95\% CI \\
    \midrule
    \shortstack{Network\\Characteristic} & Edges & 0.07$^\ddagger$ & (0.03, 0.14) & 0.09$^\ddagger$ & (0.04, 0.18) & 0.07$^\ddagger$ & (0.03, 0.14) & 0.09$^\ddagger$ & (0.04, 0.17) & 0.07$^\ddagger$ & (0.04, 0.15) & 0.09$^\ddagger$ & (0.04, 0.18) \\
    \midrule
    \multirow{3}[2]{*}{Sex} & Male - Male  & 0.61$^\dagger$ & (0.43, 0.86) & 0.52$^\ddagger$ & (0.36, 0.74) & 0.54$^\ddagger$ & (0.39, 0.75) & 0.45$^\ddagger$ & (0.32, 0.63) & 0.53$^\ddagger$ & (0.39, 0.74) & 0.44$^\ddagger$ & (0.32, 0.61) \\
          & Male - Female  & 0.59$^\dagger$ & (0.41, 0.84) & 0.51$^\ddagger$ & (0.35, 0.73) & 0.54$^\ddagger$ & (0.38, 0.75) & 0.46$^\ddagger$ & (0.33, 0.65) & 0.53$^\ddagger$ & (0.38, 0.74) & 0.46$^\ddagger$ & (0.32, 0.64) \\
          & Female - Female  & \multicolumn{12}{c}{Reference Level} \\
    \midrule
    \multirow{3}[2]{*}{Nationality} & Greek - Greek & 0.43$^\dagger$ & (0.26, 0.72) & 0.57* & (0.34, 0.96) & 0.44$^\ddagger$ & (0.27, 0.71) & 0.59* & (0.36, 0.99) & 0.44$^\ddagger$ & (0.27, 0.72) & 0.59* & (0.36, 0.99) \\
          & Greek - Not Greek & 0.34$^\ddagger$ & (0.20, 0.56) & 0.39$^\ddagger$ & (0.23, 0.67) & 0.33$^\ddagger$ & (0.20, 0.55) & 0.40$^\ddagger$ & (0.24, 0.68) & 0.34$^\ddagger$ & (0.20, 0.56) & 0.40$^\ddagger$ & (0.24, 0.68) \\
          & Not Greek - Not Greek & \multicolumn{12}{c}{Reference Level} \\
    \midrule
    \multirow{6}[2]{*}{\shortstack{Living\\Situation}} & Rent - Rent  & 0.21$^\ddagger$ & (0.12, 0.36) & 0.39$^\dagger$ & (0.21, 0.73) & 0.21$^\ddagger$ & (0.12, 0.36) & 0.37$^\dagger$ & (0.20, 0.70) & 0.20$^\ddagger$ & (0.12, 0.36) & 0.37$^\dagger$ & (0.20, 0.69) \\
          & Rent - No Rent & 0.18$^\ddagger$ & (0.13, 0.25) & 0.30$^\ddagger$ & (0.21, 0.43) & 0.18$^\ddagger$ & (0.13, 0.25) & 0.30$^\ddagger$ & (0.21, 0.42) & 0.18$^\ddagger$ & (0.13, 0.25) & 0.30$^\ddagger$ & (0.21, 0.42) \\
          & No Rent - No Rent & 0.25$^\ddagger$ & (0.19, 0.34) & 0.35$^\ddagger$ & (0.25, 0.48) & 0.26$^\ddagger$ & (0.19, 0.34) & 0.36$^\ddagger$ & (0.27, 0.49) & 0.26$^\ddagger$ & (0.19, 0.34) & 0.36$^\ddagger$ & (0.26, 0.49) \\
          & Rent - Homeless  & 0.23$^\ddagger$ & (0.16, 0.34) & 0.37$^\ddagger$ & (0.25, 0.54) & 0.24$^\ddagger$ & (0.16, 0.34) & 0.36$^\ddagger$ & (0.25, 0.52) & 0.23$^\ddagger$ & (0.16, 0.34) & 0.36$^\ddagger$ & (0.24, 0.52) \\
          & No Rent - Homeless & 0.32$^\ddagger$ & (0.25, 0.42) & 0.41$^\ddagger$ & (0.31, 0.54) & 0.34$^\ddagger$ & (0.26, 0.44) & 0.42$^\ddagger$ & (0.32, 0.55) & 0.33$^\ddagger$ & (0.25, 0.43) & 0.42$^\ddagger$ & (0.32, 0.55) \\
          & Homeless - Homeless  & \multicolumn{12}{c}{Reference Level} \\
    \midrule
    \multirow{3}[2]{*}{Education} & High School & 0.97  & (0.84, 1.11) & 0.99  & (0.85, 1.15) & 0.98  & (0.85, 1.12) & 0.98  & (0.85, 1.14) & 0.98  & (0.85, 1.12) & 0.99  & (0.85, 1.15) \\
          & Post High School & 1.35$^\dagger$ & (1.12, 1.63) & 1.36$^\dagger$ & (1.12, 1.65) & 1.35$^\dagger$ & (1.12, 1.62) & 1.38$^\ddagger$ & (1.14, 1.68) & 1.34$^\dagger$ & (1.11, 1.61) & 1.38$^\ddagger$ & (1.14, 1.68) \\
          & Less than High School & \multicolumn{12}{c}{Reference Level} \\
    \midrule
    \multirow{4}[2]{*}{Employment} & Unemployed & 1.24  & (0.99, 1.56) & 1.39$^\dagger$ & (1.09, 1.78) & 1.30* & (1.04, 1.62) & 1.46$^\dagger$ & (1.15, 1.85) & 1.31* & (1.05, 1.65) & 1.47$^\dagger$ & (1.16, 1.87) \\
          & Can't work: Health Reasons & 1.34$^\dagger$ & (1.08, 1.66) & 1.30* & (1.04, 1.63) & 1.36$^\dagger$ & (1.10, 1.67) & 1.32* & (1.06, 1.64) & 1.35$^\dagger$ & (1.09, 1.67) & 1.31* & (1.05, 1.63) \\
          & Other & 1.31* & (1.02, 1.70) & 1.31* & (1.002, 1.70) & 1.36* & (1.06, 1.75) & 1.32* & (1.02, 1.71) & {1.35*} & {(1.05, 1.73)} & 1.31* & (1.01, 1.70) \\
          & Employed & \multicolumn{12}{c}{Reference Level} \\
    \bottomrule
    \end{tabular}%
  \end{center}
  \label{tab:addlabel}%
   \flushleft \footnotesize{
OR = odds ratio\\
95\% CI is the 95\% confidence interval of the odds ratio\\
* indicates $p < 0.05$, $\dagger$ indicates $p < 0.01$, and $\ddagger$ indicates $p < 0.001$}
\end{table}%
\end{sidewaystable}
\newpage
\section{Exponential random graph models (ERGM)}
Using notation from \cite{kolaczyk2009}, based on that of \cite{pattison2007}, one defines the graph as $G=(V, E)$ where $V$ is the set of nodes (i.e., participants) and $E$ the set of edges (i.e., connection between participants), and set $Y=[Y_{ij}]$ as the random adjacency matrix to write an ERGM, which has the following exponential family form for the joint distribution of the elements in $Y$:
$$P_\theta(Y=y)=\frac{1}{k}\exp(\sum_H \theta_H\cdot g(y, x))$$
where $$g(y, x)=\sum_{1\leq i<j\leq N_\nu}y_{ij}h(x_i, x_j),$$ each $H$ is a configuration that is a set of possible edges among a subset of the vertices in the graph, $x_i$ and $x_j$ are the vector of observed attribute variables for the $i$-th and $j$-th vertices, $h(x_i, x_j)$ is a symmetric function of $x_i$ and $x_j$, $\theta_H$ represents a vector of the coefficients of $h$, and $k=k(\theta)$ is the normalizing constant
$$k(\theta)=\sum_y \exp(\sum_H \theta_H\cdot g(y, x)).$$
ERGMs use Markov chain Monte Carlo (MCMC) maximum likelihood estimation to estimate the log-odds of the effects of parameters on the observed networks following the form:
$$\log\Big[\frac{P_\theta(Y_{ij}=1|Y_{-ij}=y_{-ij}, X=x)}{P_\theta(Y_{ij}=0|Y_{-ij}=y_{-ij}, X=x)}\Big]=\theta^T\Delta_{ij}(y, x)$$
where $\Delta_{ij}(y, x)$ represents the change statistics, i.e. the difference between $g(y, x)$ when $y_{ij}=1$ and $y_{ij}=0$, $Y_{-ij}=Y\backslash\{Y_{ij}\}$, and $\theta$ is the vector of parameters being estimated. A non-zero value for $\theta$ means that the $Y_{ij}$ are dependent for all pairs of vertices $\{i, j\}$, conditional upon the rest of the graph.

These terms were included in the models as functions following the form:
$$h(x_i, x_j)=I(x_i=x_j)$$ where the indicator function, $I$, is used to indicate whether or not there is a match between two nodes’ attributes. These are also terms for second-order effects representing homophily based on attributes \citep{kolaczyk2009}. Main effects were also included in the models, following the form:
$$h(x_i, x_j)=x_i+x_j$$
these effects are additive and were included as node factor terms, main effect of a factor attribute, further explained below. 

Node match is a term that represents the odds of two people being connected to one another based on sharing a certain level of a attribute and is also generally called homophily \citep{hunter2008}. For example, for the attribute sex, we could look at the odds of two females sharing a tie (or the odds of two males sharing a tie). Node factor is used to compare people across levels of an attribute variable to see if people are more or less likely to have ties in the network, compared to a reference level \citep{hunter2008}. For example, we could see if people who are homeless are more likely to have ties than people who are not homeless, which could help researchers identify which populations might be at higher risk for engaging in risky health behaviors. Node mix is used to compare pairs of people within and between levels of an attribute variable and compares the pair’s likelihood of having ties compared to a pair reference level \citep{hunter2008}. For example, we could see if people who live in their own place are more or less likely to have ties to others who also live in their own place, compared to ties between two people who are both homeless.

\bibliography{reference}

\begin{thebibliography}{25}
\providecommand{\natexlab}[1]{#1}
\providecommand{\url}[1]{\texttt{#1}}
\expandafter\ifx\csname urlstyle\endcsname\relax
  \providecommand{\doi}[1]{doi: #1}\else
  \providecommand{\doi}{doi: \begingroup \urlstyle{rm}\Url}\fi

\bibitem[Altice et~al.(2010)Altice, Kamarulzaman, Soriano, Schechter, and
  Friedland]{altice2010}
F.~Altice, A.~Kamarulzaman, V.~Soriano, M.~Schechter, and G.~Friedland.
\newblock Treatment of medical, psychiatric, and substance-use comorbidities in
  people infected with {HIV} who use drugs.
\newblock \emph{The Lancet}, 376\penalty0 (9738):\penalty0 367--387, 2010.

\bibitem[Burnett et~al.(2018)Burnett, Broz, Spiller, Wejnert, and
  Paz-Bailey]{burnett2018}
J.~Burnett, D.~Broz, M.~Spiller, C.~Wejnert, and G.~Paz-Bailey.
\newblock {HIV} {I}nfection and {HIV-A}ssociated {B}ehaviors {A}mong {P}ersons
  {W}ho {I}nject {D}rugs - 20 {C}ities, {U}nited {S}tates, 2015.
\newblock \emph{MMWR. Morbidity and Mortality Weekly Report}, 67:\penalty0
  23--28, 01 2018.

\bibitem[{Centers for Disease Control and Prevention}(2018)]{cdc2018}
{Centers for Disease Control and Prevention}.
\newblock {HIV} {S}urveillance {R}eport, 2017; vol. 29.
\newblock \url{http://www.cdc.gov/hiv/library/reports/hiv-surveillance.html},
  11 2018.
\newblock Last Accessed March 02, 2021.

\bibitem[D'Agostino and Rubin(2000)]{dagostino2000}
R.~D'Agostino and D.~Rubin.
\newblock Estimating and {U}sing {P}ropensity {S}cores with {P}artially
  {M}issing {D}ata.
\newblock \emph{Journal of the American Statistical Association}, 95:\penalty0
  749--759, 09 2000.

\bibitem[Dombrowski et~al.(2013)Dombrowski, Khan, Mclean, Curtis, Wendel,
  Misshula, and Friedman]{dombrowski2013}
K.~Dombrowski, B.~Khan, K.~Mclean, R.~Curtis, T.~Wendel, E.~Misshula, and
  S.~Friedman.
\newblock A {R}eexamination of {C}onnectivity {T}rends via {E}xponential
  {R}andom {G}raph {M}odeling in {T}wo {IDU} {R}isk {N}etworks.
\newblock \emph{Substance Use \& Misuse}, 48\penalty0 (14):\penalty0
  1485--1497, 2013.

\bibitem[Friedman et~al.(2002)Friedman, Curtis, Neaigus, Jose, and
  Des~Jarlais]{friedman2002}
S.~Friedman, R.~Curtis, A.~Neaigus, B.~Jose, and D.~Des~Jarlais.
\newblock \emph{Social {N}etworks, {D}rug {I}njectors' {L}ives, and
  {HIV/AIDS}}.
\newblock Springer, Boston, MA, 2002.

\bibitem[Friedman et~al.(2004)Friedman, Maslow, Bolyard, Sandoval,
  Mateu-Gelabert, and Neaigus]{intravention2004}
S.~Friedman, C.~Maslow, M.~Bolyard, M.~Sandoval, P.~Mateu-Gelabert, and
  A.~Neaigus.
\newblock Urging {O}thers to be {H}ealthy: “{I}ntravention” by {I}njection
  {D}rug {U}sers as a {C}ommunity {P}revention {G}oal.
\newblock \emph{AIDS education and prevention : official publication of the
  International Society for AIDS Education}, 16:\penalty0 250--63, 07 2004.

\bibitem[Friedman et~al.(2017)Friedman, Mateu-Gelabert, Ruggles, Goodbody,
  Syckes, Jessell, Teubl, and Guarino]{friedman2017}
S.~Friedman, P.~Mateu-Gelabert, K.~Ruggles, E.~Goodbody, C.~Syckes, L.~Jessell,
  J.~Teubl, and H.~Guarino.
\newblock Sexual {R}isk and {T}ransmission {B}ehaviors, {P}artnerships and
  {S}ettings {A}mong {Y}oung {A}dult {N}onmedical {O}pioid {U}sers in {N}ew
  {Y}ork {C}ity.
\newblock \emph{AIDS and Behavior}, 21\penalty0 (4):\penalty0 994--1003, 01
  2017.

\bibitem[Gile and Handcock(2006)]{gile2006}
K.~Gile and M.~Handcock.
\newblock Model-based assessment of the impact of missing data on inference for
  networks.
\newblock 10 2006.

\bibitem[Girard et~al.(2015)Girard, Hett, and Schunk]{nodalattribute2014}
Y.~Girard, F.~Hett, and D.~Schunk.
\newblock How individual characteristics shape the structure of social
  networks.
\newblock \emph{Journal of Economic Behavior \& Organization}, 115:\penalty0
  197--216, 07 2015.

\bibitem[Hadjikou et~al.(2020)Hadjikou, Pavlopoulou, Pantavou, Georgiou,
  Williams, Christaki, Voskarides, Lavranos, Lamnisos, Pouget, Friedman, and
  Nikolopoulos]{druginjection2020}
A.~Hadjikou, I.~D. Pavlopoulou, K.~Pantavou, A.~Georgiou, L.~D. Williams,
  E.~Christaki, K.~Voskarides, G.~Lavranos, D.~Lamnisos, E.~R. Pouget, S.~R.
  Friedman, and G.~K. Nikolopoulos.
\newblock Drug {I}njection-{R}elated {N}orms and {H}igh-{R}isk {B}ehaviors of
  {P}eople {W}ho {I}nject {D}rugs in {A}thens, {G}reece.
\newblock \emph{AIDS Research and Human Retroviruses}, 11 2020.

\bibitem[Hatzakis et~al.(2015)Hatzakis, Sypsa, Paraskevis, Nikolopoulos,
  Tsiara, Micha, Panopoulos, Malliori, Psichogiou, Pharris, Wiessing, Van~de
  Laar, Donoghoe, Heckathorn, Friedman, and Des~Jarlais]{aristotle}
A.~Hatzakis, V.~Sypsa, D.~Paraskevis, G.~Nikolopoulos, C.~Tsiara, K.~Micha,
  A.~Panopoulos, M.~Malliori, M.~Psichogiou, A.~Pharris, L.~Wiessing, M.~Van~de
  Laar, M.~Donoghoe, D.~Heckathorn, S.~Friedman, and D.~Des~Jarlais.
\newblock Design and baseline findings of a large-scale rapid response to an
  {HIV} outbreak in people who inject drugs in athens, greece: the {ARISTOTLE}
  programme.
\newblock \emph{Addiction}, 110:\penalty0 1453--67, 05 2015.

\bibitem[Hunter et~al.(2008)Hunter, Goodreau, and Handcock]{hunter2008}
D.~Hunter, S.~Goodreau, and M.~Handcock.
\newblock Goodness of {F}it of {S}ocial {N}etwork {M}odels.
\newblock \emph{Journal of the American Statistical Association}, 103:\penalty0
  248--258, 02 2008.

\bibitem[Kilwein et~al.(2018)Kilwein, Hunt, and Looby]{fentanyluse2018}
T.~Kilwein, P.~Hunt, and A.~Looby.
\newblock A {D}escriptive {E}xamination of {N}onmedical {F}entanyl {U}se in the
  {U}nited {S}tates: {C}haracteristics of {U}se, {M}otives, and {C}onsequences.
\newblock \emph{Journal of Drug Issues}, 48:\penalty0 002204261876572, 04 2018.

\bibitem[Kolaczyk(2009)]{kolaczyk2009}
E.~Kolaczyk.
\newblock \emph{Statistical {A}nalysis of {N}etwork {D}ata: {M}ethods and
  {M}odels}.
\newblock Springer-Verlag New York, 2009.

\bibitem[Latkin et~al.(1995)Latkin, Mandell, Oziemkowska, Celentano, Vlahov,
  Ensminger, and Knowlton]{latkin1995}
C.~Latkin, W.~Mandell, M.~Oziemkowska, D.~Celentano, D.~Vlahov, M.~Ensminger,
  and A.~Knowlton.
\newblock Using social network analysis to study patterns of drug use among
  urban drug users at high risk for {HIV/AIDS}.
\newblock \emph{Drug and Alcohol Dependence}, 38\penalty0 (1):\penalty0 1--9,
  04 1995.

\bibitem[Little and Rubin(2002)]{little2002}
R.~Little and D.~Rubin.
\newblock \emph{Statistical analysis with missing data. 2nd edn.}
\newblock John Wiley \& Sons, Inc, 2002.

\bibitem[Malina et~al.(2015)Malina, Frieden, Foti, and Mermin]{frieden2015}
D.~Malina, T.~Frieden, K.~Foti, and J.~Mermin.
\newblock Applying {P}ublic {H}ealth {P}rinciples to the {HIV} epidemic —
  {H}ow {A}re {W}e {D}oing?
\newblock \emph{New England Journal of Medicine}, 373\penalty0 (23):\penalty0
  2281--2287, 12 2015.

\bibitem[Nikolopoulos et~al.(2016)Nikolopoulos, Pavlitina, Muth, Schneider,
  Psichogiou, Williams, Paraskevis, Sypsa, Magiorkinis, Smyrnov, Korobchuk,
  Vasylyeva, Skaathun, Malliori, Kafetzopoulos, Hatzakis, and
  Friedman]{nikolopoulos2016}
G.~Nikolopoulos, E.~Pavlitina, S.~Muth, J.~Schneider, M.~Psichogiou,
  L.~Williams, D.~Paraskevis, V.~Sypsa, G.~Magiorkinis, P.~Smyrnov,
  A.~Korobchuk, T.~Vasylyeva, B.~Skaathun, M.~Malliori, E.~Kafetzopoulos,
  A.~Hatzakis, and S.~Friedman.
\newblock A network intervention that locates and intervenes with recently
  {HIV}-infected persons: The transmission reduction intervention project
  {(TRIP)}.
\newblock \emph{Scientific Reports}, 6:\penalty0 38100, 12 2016.

\bibitem[Robins et~al.(2007)Robins, Pattison, Kalish, and Lusher]{pattison2007}
G.~Robins, P.~Pattison, Y.~Kalish, and D.~Lusher.
\newblock An {I}ntroduction to {E}xponential {R}andom {G}raph (p*) {M}odels for
  {S}ocial {N}etworks.
\newblock \emph{Social Networks}, 29:\penalty0 173--191, 05 2007.

\bibitem[Selik and Linley(2018)]{selik2018}
R.~Selik and L.~Linley.
\newblock Viral loads within 6 weeks after diagnosis of {HIV} infection in
  early and later stages ({P}reprint).
\newblock \emph{JMIR Public Health and Surveillance}, 4:\penalty0 e10770, 04
  2018.

\bibitem[Stekhoven and Bühlmann(2012)]{stekhoven2012}
D.~Stekhoven and P.~Bühlmann.
\newblock Miss{F}orest? {N}on-parametric missing value imputation for
  mixed-type data.
\newblock \emph{Bioinformatics (Oxford, England)}, 28:\penalty0 112--8, 01
  2012.

\bibitem[Unger et~al.(2006)Unger, Kipke, Rosa, Hyde, Ritt-Olson, and
  Montgomery]{unger2006}
J.~Unger, M.~Kipke, C.~Rosa, J.~Hyde, A.~Ritt-Olson, and S.~Montgomery.
\newblock Needle-sharing among young {IV} drug users and their social network
  members: The influence of the injection partner's characteristics on {HIV}
  risk behavior.
\newblock \emph{Addictive behaviors}, 31:\penalty0 1607--18, 10 2006.

\bibitem[Volz et~al.(2013)Volz, Ionides, Romero-Severson, Brandt, Mokotoff, and
  Koopman]{volz2013}
E.~Volz, E.~Ionides, E.~Romero-Severson, M.-G. Brandt, E.~Mokotoff, and
  J.~Koopman.
\newblock {HIV-1} {T}ransmission during {E}arly {I}nfection in {M}en {W}ho
  {H}ave {S}ex with {M}en: {A} {P}hylodynamic {A}nalysis.
\newblock \emph{PLoS medicine}, 10:\penalty0 e1001568, 12 2013.

\bibitem[Ware et~al.(1981)Ware, Brook, Davies, and Lohr]{ware1981}
J.~Ware, R.~Brook, A.~Davies, and K.~Lohr.
\newblock Choosing measures of health status for individuals in general
  population.
\newblock \emph{American journal of public health}, 71:\penalty0 620--5, 07
  1981.

\end{thebibliography}

\end{document}